\documentclass[twocolumn]{aastex631}

\newcommand\degree{{^\circ}}

\newcommand\pc{{\rm\,pc}}

\newcommand\kpc{{\rm\,kpc}}

\newcommand\Gyr{{\rm\,Gyr}}

\newcommand\kmsec{{\rm\,km\,s^{-1}}}
\newcommand\kms{\kmsec}

\newcommand\s{{\rm\,s}}
\newcommand\K{{\rm\,K}}

\newcommand\mags{{\rm\,mag}}
\newcommand\medAV{$\widetilde{A_V}$}
\newcommand\fred{$f_{red}$}

\received{March 8, 2023}
\revised{April 10, 2023}
\accepted{\today}

\submitjournal{ApJ}

\shortauthors{Dalcanton et al.}
\shorttitle{M31's PHAT Stellar Disk}

\graphicspath{{./}{}}

\begin{document}

\title{The Panchromatic Hubble Andromeda Treasury XX: The Disk of M31 is Thick}

\correspondingauthor{Julianne Dalcanton}
\email{jdalcanton@flatironinstitute.org}

\author[0000-0002-1264-2066]{Julianne J.\ Dalcanton}
\affiliation{Center for Computational Astrophysics, Flatiron Institute, 162 Fifth Ave, New York, NY 10010, USA}
\affiliation{Department of Astronomy, Box 351580, University of
  Washington, Seattle, WA 98195}

\author[0000-0002-5564-9873]{Eric F.~Bell}
\affiliation{Department of Astronomy, University of Michigan, 1085 S.\ University Ave., Ann Arbor, MI 48109, USA}

\author[0000-0003-1680-1884]{Yumi Choi}
\affiliation{Department of Astronomy, University of California Berkeley, Berkeley, CA 94720, USA}

\author[0000-0001-8416-4093]{Andrew E. Dolphin}
\affiliation{Raytheon, 1151 E. Hermans Road, Tucson, AZ 85706}

\author[0000-0001-9256-5516]{Morgan Fouesneau}
\affiliation{Max Planck Institute f\"{u}r Astronomie, K\"{o}nigstuhl 17, 69117, Heidelberg, Germany}

\author[0000-0002-6301-3269]{L\'eo Girardi}
\affiliation{Osservatorio Astronomico di Padova -- INAF, 
  Vicolo dell'Osservatorio 5, I-35122 Padova, Italy}

\author[0000-0003-2866-9403]{David W. Hogg}
\affiliation{Center for Cosmology and Particle Physics, Department of Physics, New York University, 4 Washington Pl \#424, New York, NY, 10003 USA}
\affiliation{Center for Computational Astrophysics, Flatiron Institute, 162 Fifth Ave, New York, NY 10010, USA}

\author[0000-0003-0248-5470]{Anil C.~Seth}
\affiliation{University of Utah, Salt Lake City, UT, USA}

\author[0000-0002-7502-0597]{Benjamin F.~Williams}
\affiliation{Department of Astronomy, Box 351580, University of
  Washington, Seattle, WA 98195}

\begin{abstract}

We present a new approach to measuring the thickness of a partially face-on stellar disk, using dust geometry. In a moderately-inclined disk galaxy, the fraction of reddened stars is expected to be 50\% everywhere, assuming that dust lies in a thin midplane.  In a thickened disk, however, a wide range of radii project onto the line of sight. Assuming stellar density declines with radius, this geometrical projection leads to differences in the numbers of stars on the near and far sides of the thin dust layer. The fraction of reddened stars will thus differ from the 50\% prediction, with a deviation that becomes larger for puffier disks.  We map the fraction of reddened red giant branch (RGB) stars across M31, which shows prominent dust lanes on only one side of the major axis. The fraction of reddened stars varies systematically from 20\% to 80\%, which requires that these stars have an exponential scale height $h_z$ that is $0.14\pm0.015$ times the exponential scale length ($h_r\approx5.5\kpc$). M31's RGB stars must therefore have $h_z=770\pm80\pc$, which is far thicker than the Milky Way's thin disk, but comparable to its thick disk. The lack of a significant thin disk in M31 is unexpected, but consistent with its interaction history and high disk velocity dispersion. We suggest that asymmetric reddening be used as a generic criteria for identifying ``thick disk'' dominated systems, and discuss prospects for future 3-dimensional tomographic mapping of the gas and stars in M31.

\end{abstract}


\section{Introduction} \label{introsec}

Historically, the Milky Way has set our understanding of the structure
of galaxy disks.  The classical picture of massive disks --- a
rapidly-rotating thin stellar disk, embedded in a less massive,
thicker, slowly-rotating stellar disk, both surrounded by an even more
diffuse stellar halo of fast moving stars --- all developed in
response to studies of the structure and kinematics of Milky Way stars
\citep[see the review by][]{blandhawthorn2016}.  

Over the years, evidence has accumulated that many of these same
features are present in other massive galaxies.  Thickened, rotating
stellar envelopes appear to be common around edge-on galaxies
\citep[e.g.,][]{dalcanton2002, yoachim2008, comeron2011,
  elmegreen2017}, and more extended stellar halos have been revealed in
many galaxies \citep[e.g., ][]{mouhcine2010,ibata2014,greggio2014,
  rejkuba2014,crnojevic2016,merritt2016,trujillo2016,harmsen2017,dsouza2018,
  wu2021,beraldo2021,gilbert2022},
though whether they are better analogs of the Milky Way's low
metallicity hot stellar halo or its tidal debris is often
unclear.

Within the Milky Way, the thin$+$thick disk and halo ansatz has become
necessarily more complex as better data has accumulated and has moved
beyond star counts to include kinematics and metal abundances (see,
for example, reviews by \citet{blandhawthorn2016} and \citet{helmi2020}, noting the added complexity introduced by the identification of the Gaia-Encedelus "sausage" \citep[e.g.,][]{helmi2018,haywood2018,belokurov2018,deason2018}).
However, the
basic utility of the original picture has remained a useful framework
outside of the bulge/bar-dominated inner regions, given that it
broadly separates a galaxy into a ``high specific angular momentum,
low velocity dispersion'' component, a ``high specific angular
momentum, moderate velocity dispersion'' component, and ``high
velocity dispersion, low net angular momentum\footnote{ Individual
  tidal streams that contribute to the halo may have high specific
  angular momentum, but having an extended, roughly spherical
  distribution suggests that the net specific angular momentum is not
  high}'' component.  It is particularly hard to justify more nuanced
decompositions of external galaxies where the data are not as rich,
and it becomes impossible to decompose unresolved, low-density
stellar populations by their detailed kinematics, photometry, and abundance patters, and
where all of the structural and kinematic features can only be seen in
projection.

This latter limitation is especially vexing, particularly when
attempting to separate thick and thin disk subcomponents.  When a
galaxy is seen edge-on, one can potentially decompose the disk into 2
components, albeit with substantial uncertainties in inferring the
thin disk structure due to dust. However, there is no way to
simultaneously measure the vertical kinematics of either
component. The opposite problem occurs when a galaxy is more
face-on. At best, one might be able to detect the signature of a thick
disk in the kinematics of individual stars, but one could not actually
measure its height directly. In short, simultaneously measuring the
structure and velocity dispersions of thin and thick disk components
becomes essentially intractable outside the Milky Way.

In this paper, we take advantage of a novel probe of disk structure
along the line of sight that breaks the degeneracy, allowing us to
solve for the vertical structure in M31 --- a face-on but partially
inclined galaxy. Using the dust mapping technique from
\citet{dalcanton2015}, we measure the fraction of stars that lie
behind M31's layer of dusty ISM.  We then show how the amplitude and
spatial variation in this ``reddening fraction'' can be used to
constrain the inclination of M31 and the thickness of its stellar
disk.  We show that M31's stellar disk must be much thicker than that
of the Milky Way, and as such is consistent with M31's complex halo
structure \citep[e.g.,][]{ibata2014}, high internal stellar
velocity dispersion \citep[e.g.,][]{dorman2015}, recent merger-driven
burst of star formation \citep{williams2015,bernard2015}, and weak
planetary nebula metallicity gradient \citep[e.g.,][]{jacoby1999,kwitter2012,
  balick2013, balick2017}, which is likely due to projection effects smoothing out an
  intrinsically steeper population gradient in the older stars.

Beyond the measurement of M31's disk structure, we discuss further possibilities for using the fraction of reddened stars as a diagnostic of stellar and ISM geometry. The model calculations presented here are quite general, and point to using apparent asymmetries in the reddening across galaxies as a generic signpost of significantly thickened stellar disks, even when the sort of detailed analysis in \citet{dalcanton2015} is not available.  Given that the thickness of stellar disks has long been recognized as an indicator of past dynamical heating from interactions (e.g., going back to \citet{toth1992}), a census of disk thickness that includes galaxies that are not fully edge-on would be illuminating.  We also discuss the wealth of information that can be extracted from local departures from the smooth models presented here.  Any warping of the gas or stars (globally, or, with respect of one to the other) will lead to deviations from the expected reddening fraction \citep[e.g.,][]{choi2018, yanchulova2021}, as would any significant offset of the dusty ISM from the stellar midplane, such as might be expected for infalling or accreted gas.  These features turn maps of dust reddening fractions into powerful constraints on the three-dimensional distribution of the dense ISM and of warps in both the gas and stars.

\section{Measuring Disk Geometry Using Dust} \label{overviewsec}

When the dusty ISM is in a thin layer, it is reasonable to assume that
most stars will lie either in front of or behind the dust, with only a
small fraction of stars being embedded within.  With this
assumption, which is most likely to hold for older, vertically-heated
stellar populations, roughly half of a galaxy's stars will be behind
the dusty gas when viewed face-on, provided that the disk is
sufficiently undisturbed that the gas has settled into the midplane.

\begin{figure*}
\centerline{
\includegraphics[width=6.5in]{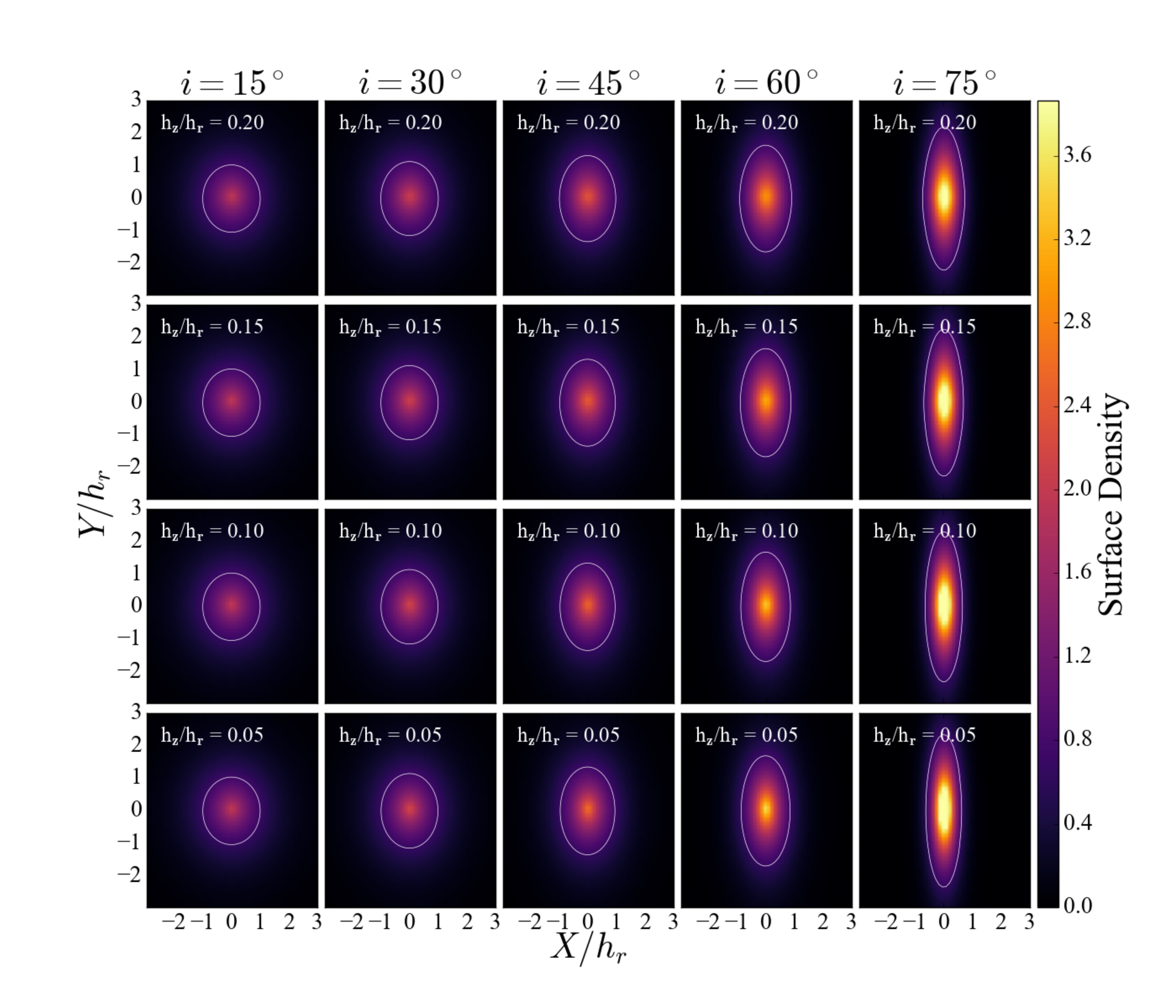}
}
\caption{Maps of the apparent surface density for models of inclined,
  thickened disks, normalized to have a central surface density of
  unity when viewed face-on. The apparent inclinations of the model
  disks increase from left to right ($i=15\degree$, $30\degree$,
  $45\degree$, $60\degree$, \& $75\degree$), and the disk thickness
  increases from bottom to top ($h_z/h_r=0.05$, 0.1, 0.15, \& 0.2).
  The white ellipse is drawn at a constant apparent surface density,
  equal to that seen at $r=h_r$ for a face-on disk.  At large
  inclination, this isodensity contour moves to larger effective
  radii, due to the increased path length through the disk.
\label{surfdensfig}}
\end{figure*}

When the same galaxy is viewed inclined along the line of sight,
however, the apparent fraction of reddened stars can deviate from the
expected value of one-half, even though the disk$+$dust geometry has
remained fixed.  Unlike a face-on disk, a given line of sight through
an inclined disk samples a range of galactic radii.  Because the stellar
surface density drops with radius, stars in the inner disk will be
overrepresented along a given line of sight.  If those inner disk
stars are on the near side of the dust, then the fraction of reddened
stars will fall below one-half along that particular line of
sight. Conversely, if those inner disk stars are on the far side, then
the apparent reddened fraction will be higher than one-half.  This
effect of geometry will therefore imprint a spatial pattern on a map
of the fraction of reddened stars, depending on whether one is viewing
the near or far side of the inclined disk.  \citet{elmegreen1999} have
previously used this effect to explain why asymmetric patterns of dust
reddening do not necessarily imply asymmetric distributions of the
dust itself, and in M31 \citet{merrett2006} have used this effect to
explain spatial variations in the planetary nebula luminosity
function.

The size of this effect depends strongly on position within the
galaxy, and on the disk$+$dust geometry.  The range of radii that a
given line of sight samples (and thus the amplitude of the shift in
reddening fraction) will be larger when the disk is intrinsically
thicker, or is more inclined along the line of sight.  In addition, the
effect will be negligible along the major axis, where lines of sight
sample a range of azimuthal angles, but all at approximately the same
radii for an undisturbed disk. The major axis will therefore always 
show a reddened fraction
of 50\% assuming it is aligned with the line of nodes.

The net result is that a map of the fraction of reddened stars is an
excellent probe of disk structure and viewing geometry. The location
of the 50\% reddening line indicates the location of the major axis at
each radius, and is therefore sensitive to the position angle and its
radial variation. The rapidity with which the reddening fraction
varies with distance from the major axis, and the amplitude of that
variation, simultaneously constrains the inclination of the disk and
its thickness compared to the radial scale length of the disk.  Since
the radial scale length can be measured along the major axis itself,
modeling the map of the reddening fraction therefore provides a new
way to measure the thickness of the stellar disk.

\subsection{Calculating the Positional Dependence of \fred}

We calculate the above effects by adopting a simple model where
the stellar disk is radially exponential with a scale length $h_r$ and
has a vertical exponential distribution with scale height $h_z$.  For
this model, the space density of stars as a function of radius $r$ and
height $z$ above the midplane is

\begin{equation} \label{rhoeqn}
  \rho(r,z) = \rho_0 \, e^{-r/h_r} \, e^{-\left| z \right| / h_z},
\end{equation}

\noindent where $\rho_0$ is the density in the very center of the
galaxy. We calculate the surface density of stars that are in
front of ($\Sigma_+$) or behind ($\Sigma_{-}$) the midplane by
integrating the density $\rho$ along the line of sight from the
midplane, to positive or negative infinity, respectively. The total
stellar surface density $\Sigma$ will then be $\Sigma_+ + \Sigma_{-}$,
which is equal to $\Sigma_0=2\rho_0 h_z$ in the center of the galaxy.

We perform the path integration along a variable $l$ defined to be
zero at the midplane. Along the integration path, for a galaxy with an
inclination of $i$, the height above the midplane is $z=l\,\cos{i}$ and
the radius is $r^2 = (x_0 + l\,\sin{i})^2 + y_0^2$, where ($x_0$,
$y_0$) is the cartesian coordinate on the uninclined disk, assuming
the major axis is oriented along the $y$-axis.

\begin{figure}
\centerline{
  \includegraphics[width=3.95in]{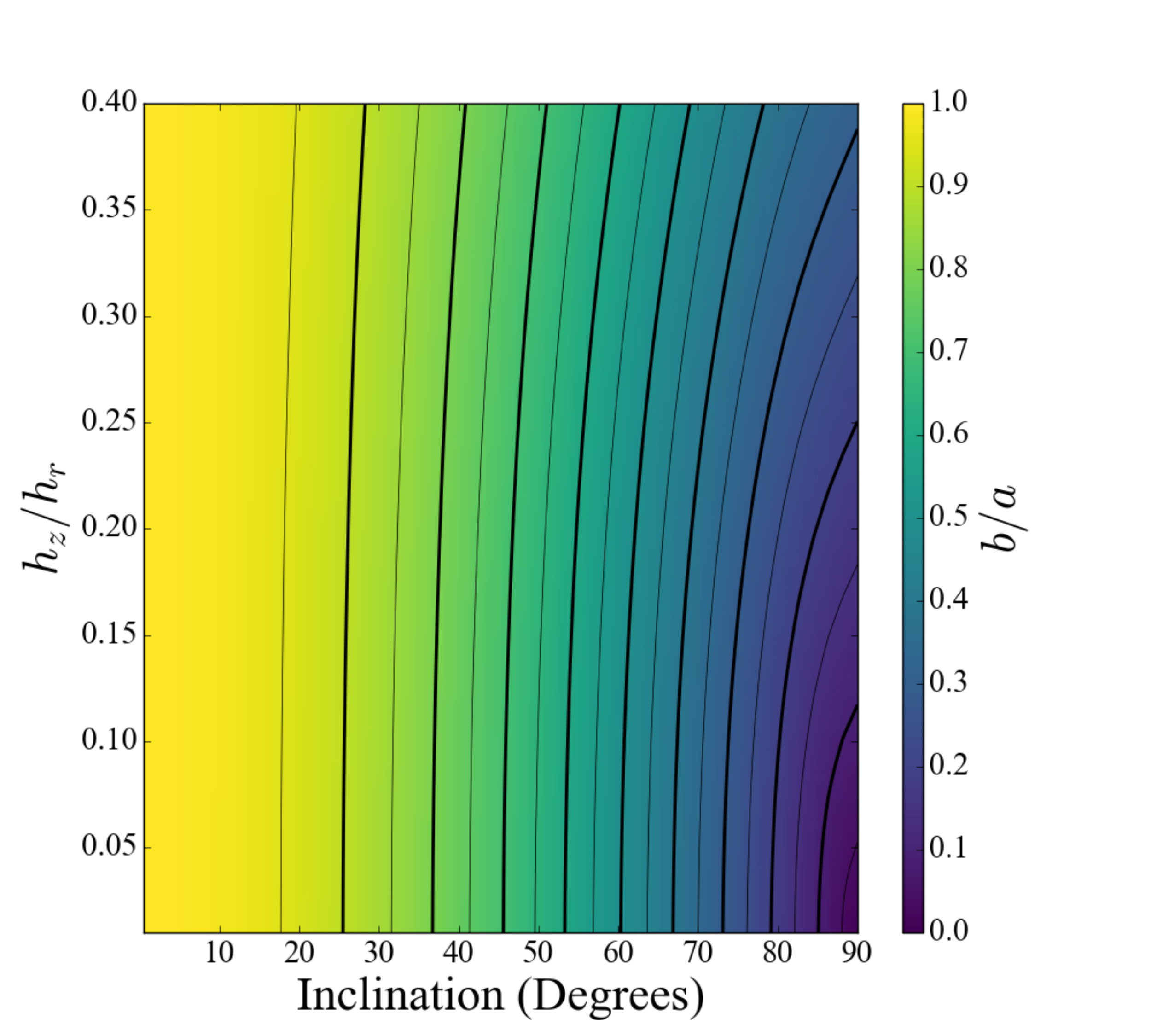}
}
\caption{The axial ratio of model inclined, thickened disks, measured
  at the surface brightness corresponding to $r=1.5h_r$ for a face-on
  disk. The heavy solid lines indicate the value of $b/a$ in steps of
  0.1 (i.e., $b/a=0.1$, 0.2, etc). The light lines indicate
  $b/a=0.05$, 0.15, etc.  For thicker disks and higher inclinations,
  the apparent axial ratio departs significantly from the naive
  expectation for an inclined infinitely thin disk.
\label{axialratiofig}}
\end{figure}

In Figure~\ref{surfdensfig} we show maps of the projected stellar
surface density, in a grid of inclination (varying along rows) and of
disk thickness (i.e., $h_z/h_r$, varying along columns, with intrinsically
thicker disks plotted towards the top). As expected, disks appear
thinner when they are more highly inclined. This variation with
inclination is less pronounced for intrinsically thicker disks,
however, because such disks can never look truly thin, even at the
highest inclinations.  The impact of inclination and disk thickness on
the apparent disk structure is summarized in
Figure~\ref{axialratiofig}, where we plot the apparent axis ratio as a
function of inclination and $h_z/h_r$. The axis ratio is
calculated for a fixed characteristic surface brightness level, chosen
to be that observed for a face-on disk at a radius of $1.5h_r$, where the surface mass density of an exponential disk would have fallen by roughly a factor of 5 from the center but still be reliably measured in survey data.  The typical radius where this is measured encompasses close to half of the mass of the disk, but is not so far out that warps and asymmetries are potentially significant.  The locii in Figure~\ref{axialratiofig} also assume that the disk is optically thin at the radius where $b/a$ was measured, which is not a bad assumption for long-wavelength observations at large radii.

\begin{figure*}
\centerline{
\includegraphics[width=6.5in]{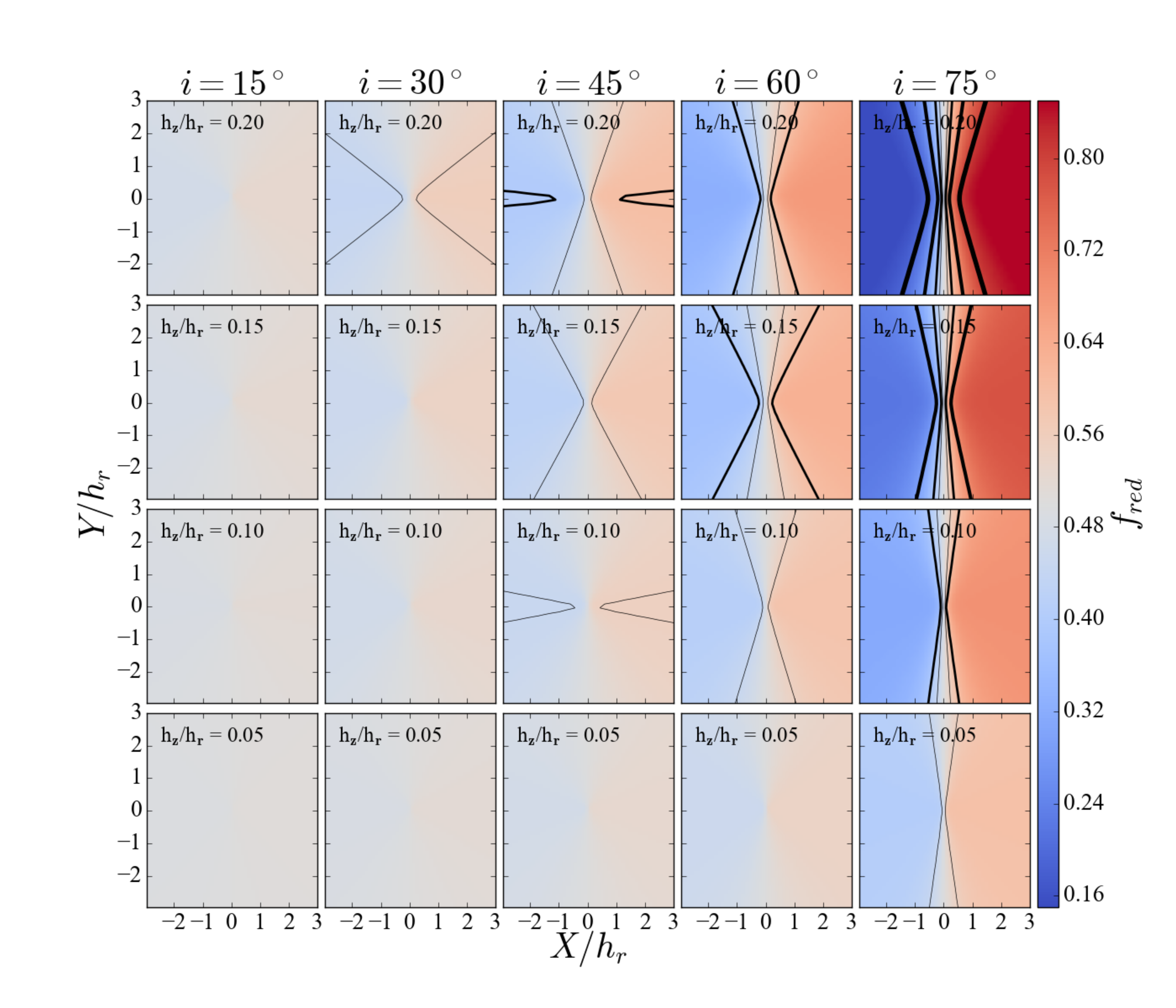}
}
\caption{Maps of the apparent fraction of reddened stars for models of
  inclined, thickened disks where the dust is assumed to be confined
  to the midplane with negligible thickness compared to the stars
  (i.e., $h_z >> h_{dust}$). The apparent inclinations of the model
  disks increases from left to right ($i=15\degree$, $30\degree$,
  $45\degree$, $60\degree$, \& $75\degree$), and the disk thickness
  increases from bottom to top ($h_z/h_r=0.05$, 0.1, 0.15, \& 0.2). As
  expected, the fraction of reddened stars is always 50\% along the
  major axis, but deviates strongly perpendicularly, with the largest
  deviations seen for higher inclinations and intrinsically thicker
  disks.  Contours indicate deviations of $\pm5$\%, $\pm10$\%,
  $\pm20$\%, \& $\pm30$\% relative to 50\%, with thicker contours
  indicating larger deviations.
\label{fredgridfig}}
\end{figure*}

Of greater interest is Figure~\ref{fredgridfig}, where we plot
maps of the fraction of reddened stars ($f_{red} \equiv
\Sigma_{-}/(\Sigma_{+}+\Sigma_{-})$), for model disks with the same
inclinations and thicknesses as in Figure~\ref{surfdensfig}. At low
inclinations, the fraction of stars behind the dust layer is
essentially constant at $f_{red}=0.5$.  At high inclinations, however,
the impact of the thickness of the disk can be seen. Off of the major
axis, lines of sight pierce a range of radii, leading to significant
differences in the fraction of reddened stars seen on either side of
the major axis. In the half of the galaxy that is furthest from the
observer, the stars in front of the dust layer come from the inner
galaxy where the number density of stars is higher, leading to low
reddening fractions.  The variation in reddening fraction
across a disk is largest for high inclinations and for intrinsically thicker disks, both of
which lead to longer path lengths and thus larger ranges of radii
along a given line of sight.

\begin{figure}
\centerline{
\includegraphics[width=3.95in]{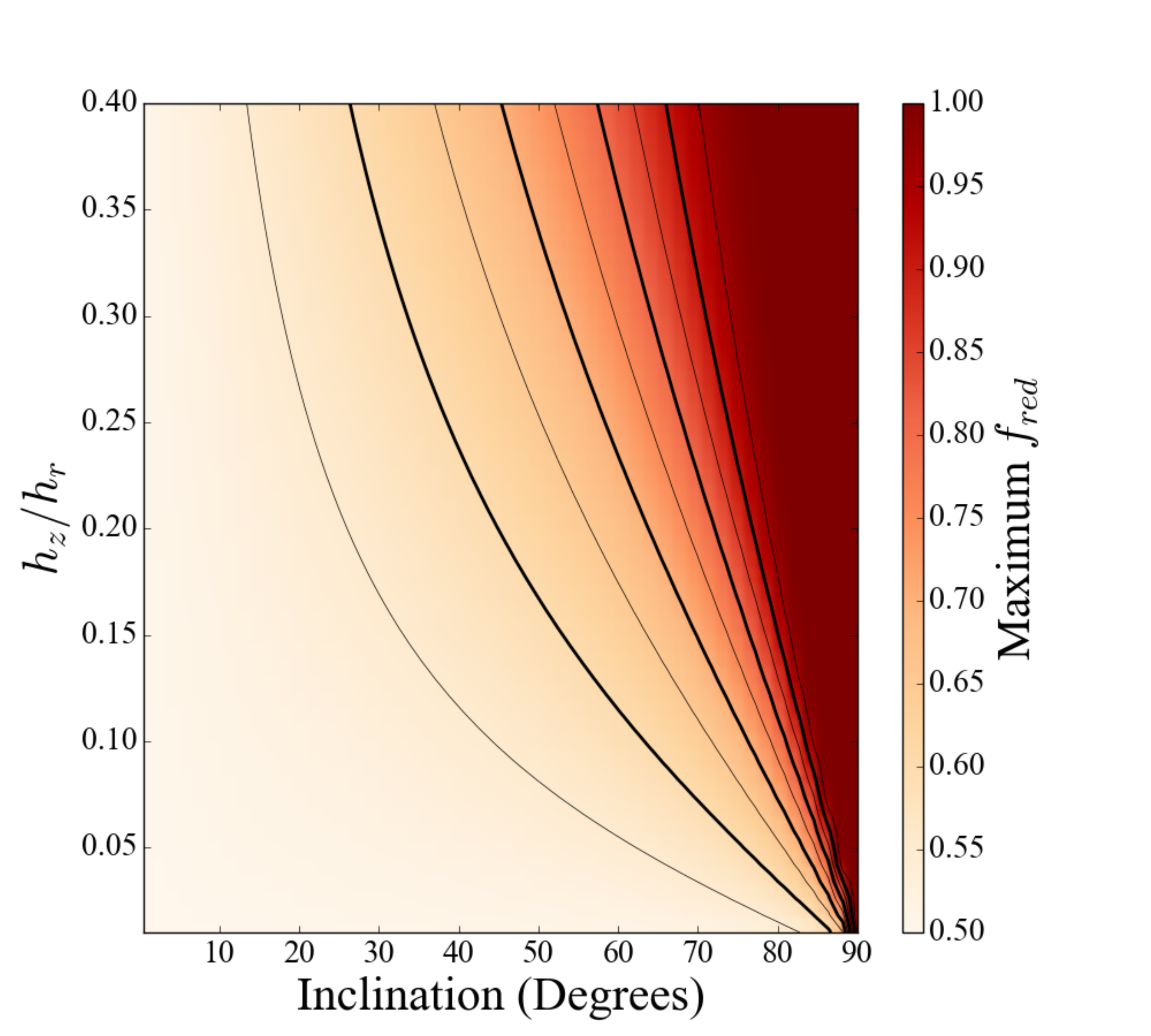}
}
\caption{The maximum reddening fraction $f_{red}$ for model inclined,
  thickened disks, measured along the minor axis at the surface
  brightness corresponding to $r=3h_r$ for a face-on disk. The heavy
  solid lines indicate the value of $f_{red}$ in steps of 0.1 (i.e.,
  $f_{red}=0.6$, 0.7, etc). The light lines indicate $f_{red}=0.55$,
  0.65, etc.  For thicker disks and higher inclinations, the maximum
  observed value of $f_{red}$ deviates most strongly from the naive
  value of $f_{red}=0.5$.  The minimum value of $f_{red}$ at each
  point is 1 minus the plotted value.
\label{fredmaxfig}}
\end{figure}

We show the full variation in reddening fraction in
Figure~\ref{fredmaxfig}, where we plot the approximate maximum value
of $f_{red}$ along the minor axis, calculated where the surface
brightness falls to a value equal to that found at $3h_r$ for a
face-on disk, which is far enough out that $f_{red}$ can be assumed to be close to the maximum, while also expecting to be potentially measurable ($\sim\!25$ mag/arcsec$^2$ for an exponential Freeman disk).  As expected from Figure~\ref{fredgridfig}, the maximum
observed value of $f_{red}$ will be higher for more inclined and/or
intrinsically thicker galaxies.

The above models point to ways in which the observed reddening
fraction can be used to make new contraints on disk geometry.  First,
in any inclined galaxy, seeing obvious dust obscuration on only one side of
the major axis trivially implies a thick stellar disk. Secondly,
comparing Figures~\ref{axialratiofig} and \ref{fredmaxfig} suggests
that the combination of measuring a disk's apparent axial ratio and its maximum
(or minimum) fraction of reddened stars will jointly constrain the disk's
thickness (through $h_z/h_r$) and inclination.  We will take
this approach below, after presenting
measurements of $f_{red}$ in M31.

\section{Measuring the Fraction of Reddened Stars in M31} \label{measurementsec}

\begin{figure*}
\centerline{
\includegraphics[width=6.5in]{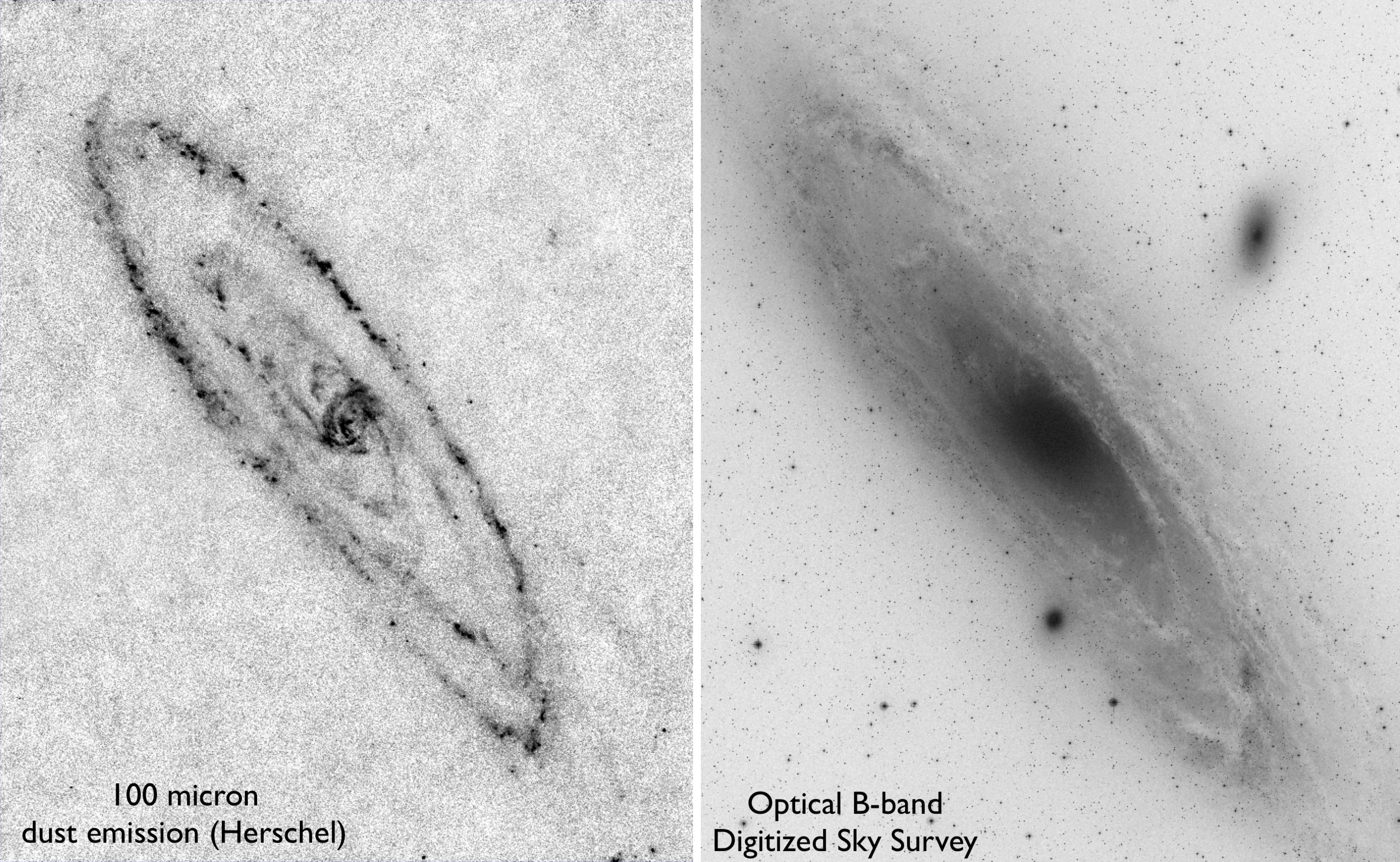}
}
\caption{Comparison between a Herschel PACS 100$\mu$m image of emission from M31's dust (left; from \citet{fritz2012}), and an optical B-band image from the Digitized Sky Survey (right). The Herschel image shows that M31's dust content is similar on either side of the major axis.  However, the stellar extinction from the dust is far more apparent on the northern (upper) half of the galaxy, leading to prominent dust lanes in the optical image.  The PHAT survey footprint roughly covers the upper left quadrant of the image, extending further out in radius beyond the prominent 10$\kpc$ ring seen in the Herschel image.  
\label{fir_optical_fig}}
\end{figure*}

M31 is a massive, inclined, Sb galaxy, whose significant bulge and relatively low star formation rate places it in the ``green valley'' between active and quiescent galaxies \citep{mutch2011}.  Its star formation is driven by a significant, structured ISM that produces visible dust lanes, as shown in Figure~\ref{fir_optical_fig}.  The left panel shows 100$\mu$m dust emission observed with Herschel \citep{fritz2012}\footnote{Accessed from https://irsa.ipac.caltech.edu/data/Herschel/HELGA/index.html}, covering out to slightly beyond M31's star forming ring at 10$\kpc$ radius\footnote{Note that there is more dust at larger radius, but it is below the sky level of this image. Filtering in the reduction pipeline also misses some of the more extended emission, particularly at large radii \citet{clark2021}.}. M31's dust content is similar on either side of the major axis, although there are some modest differences between the northeastern (upper left) and southwestern (lower right) halves of the galaxy, largely due to the additional split in the ring in the southeast.  

In contrast to the dust emission on the left, the blue optical image on the right shows a very different degree of asymmetry across the major axis. On the right (near) side of the galaxy, the dust lanes are very strong, even in regions where the absolute amount of dust is comparatively low.  The left (far) side of the galaxy, however, shows very little obvious dust extinction away from the major axis, in spite of there being ample dust.  This morphology alone suggests that M31 has an internal geometry and viewing angle comparable to the models in the upper right quadrant of Figure~\ref{fredgridfig}.

\begin{figure*}
\centerline{
\includegraphics[width=6.5in]{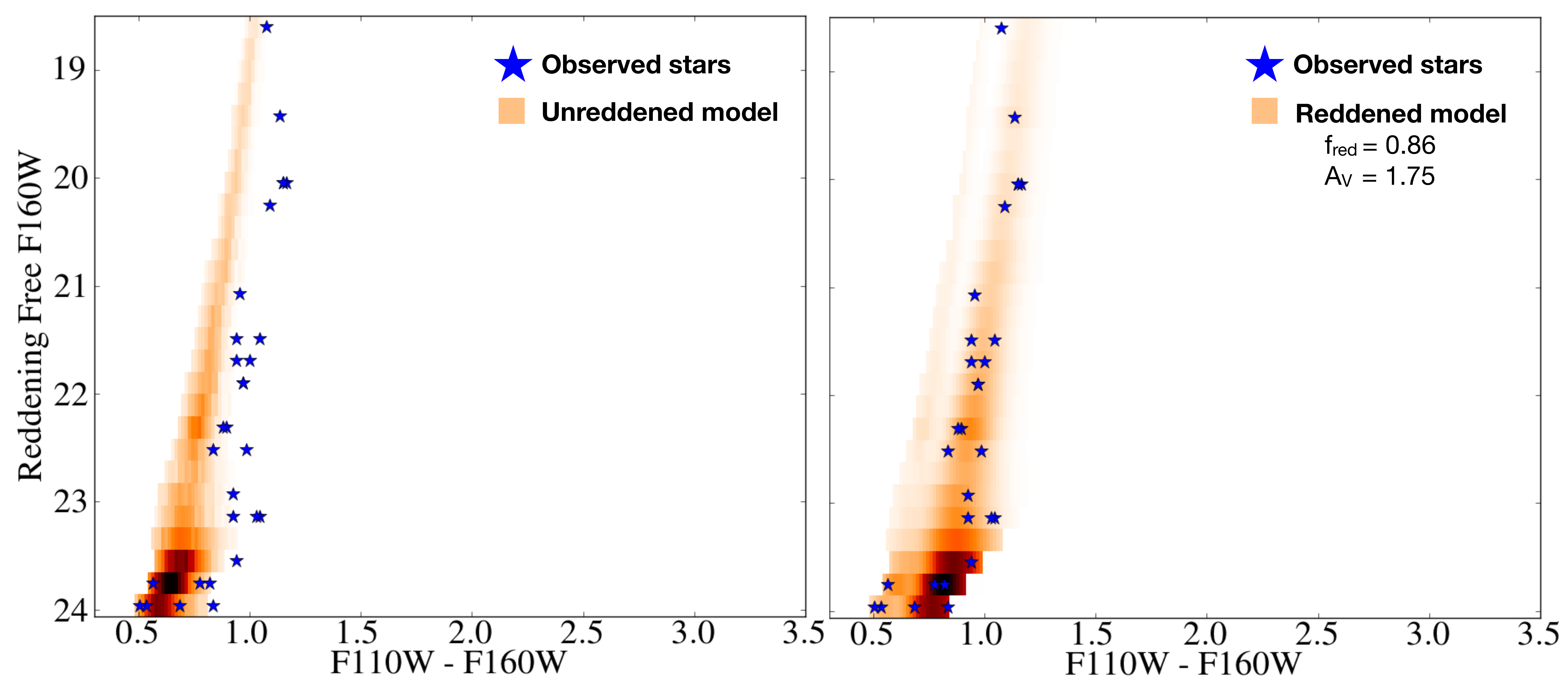}
}
\caption{A representative comparison between the observed NIR stellar photometry of the red giant branch in a single analysis pixel (blue stars), to the expected model of the unreddened RGB (left) and to the best fit model where a fraction $f_{red}$ of the RGB stars are allowed to be behind a layer of dust with a log-normal distribution of extinction characterized by a median $A_V$ (right).  This region, drawn from the northern half of the major axis, requires a large fraction of the stars to be reddened, consistent with the visual presence of dust lanes in Figure~\ref{fir_optical_fig}.  The F160W magnitude has been transformed to a reddening-free quantity, such that dust extinction and reddening move stars only to the right \citep{dalcanton2015}.
\label{model_cmd_fig}}
\end{figure*}

We can quantify the position-dependent reddening in Figure~\ref{fir_optical_fig} by measuring the fraction of reddened stars as a function of position
using the near-infrared (NIR) color-magnitude diagram (CMD) fitting
technique described in \citet{dalcanton2015}. Briefly, we subdivide
photometric data from the PHAT survey\footnote{All the {\it HST} data used in this paper can be found in MAST: \dataset[10.17909/T91S30]{http://dx.doi.org/10.17909/T91S30}}
\citep{dalcanton2012,williams2014} into $\sim6.6\arcsec$ pixels
($25\pc$ at the distance of M31). Within each pixel, we model the red
giant branch (RGB) stars in the F110W-F160W CMD as a combination of an unreddened
foreground and a reddened background population. We assume that the
reddened stars have passed through a region with a log-normal
distribution of dust columns, and then fit for the median extinction
\medAV, the width of the log-normal $\sigma$, and the fraction \fred\
of stars that are in the reddened component. Figure~\ref{model_cmd_fig} shows an example comparing the measured stellar photometry to a model of the unreddened RGB (left) and to the best fit model that contains both unreddened and reddened stars (right).

We characterize the posterior probability distribution function of
each of these parameters using a Monte Carlo Markov-Chain sampler
\citep[{\tt emcee};][]{foremanmackey2013}, assuming reasonable Baysian
prior probability distributions for \fred\ and $\sigma$.  As is
typical with Bayesian priors, the choice of the prior probability
distribution only affects the fitted values of a parameter in regions
where the constraints provided by the data are weak.  Outside of these
regions, the choice of prior has a negligible effect on the value of
the fitted parameters.

The prior for \fred\ was assigned iteratively, as follows.  We
initially fit for \fred\ with a single prior to identify regions where
the values of \fred\ were well-constrained.  We initially ran the
fitting routine with a Gaussian prior centered at $f_{red}=0.4$; we
chose this peak to be less than 0.5 because the majority of the PHAT
survey area was located on the side of M31 with less obvious dust
reddening.  This choice of a single, spatially-uniform prior is
obviously inadequate (given the maps in Figure~\ref{fredgridfig}), but
was sufficient to identify the regions where \fred\ was well-measured,
and then to use those regions to build a more appropriate
spatially-variable prior for \fred.  As discussed in
\citet{dalcanton2015}, the uncertainties in \fred\ are lowest where
the median extinction \medAV\ is highest. In these regions, the
reddened RGB is cleanly separated from the undreddened RGB on the CMD,
making the value of \fred\ unambiguous.  We therefore took all regions
with \medAV$>1$, and then fit a tilted disk model to derive a
positionally-dependent mean value of \fred.  We then solved for the
dust map parameters again using a prior that matched the expected
value of \fred\ at each location. As expected, this change in prior
had no noticeable effect on the values of \fred\ in high extinction
regions.  In the analysis that follows, we include only high extinction
pixels ($A_V>1.25$), where the prior has minimal impact.  Further
details of these procedures and their associated uncertainties can be
found in \citet{dalcanton2015}.

Throughout the fitting, we implicitly assume that stars are either in
front of or behind the dusty gas.  While this assumption is unlikely
to hold for young stars forming out of the gas, it is likely to be
valid for the older stars that dominate the red giant branch.
Empirically, we know that typical massive disk galaxies have clear
dust lanes when viewed edge-on, strongly suggesting that the dusty,
cold ISM is found in a layer that is much thinner than that of the
stars.  Quantitatively, in our own galaxy, the stellar disk has an
exponential scale height $h_{z,stars}$ of $300\pc$ and $900\pc$, for
the thin and thick disks, respectively \citep{juric2008}.  These
heights can be compared to the scale height of the cold dust ISM, for
which CO observations find a much smaller vertical half-width half-max
of $z_{hwhm} \!\sim\!50\pc$ (or
$h_{z,dust}=1.4z_{HWHM}\approx72\pc$) within the solar circle, with
likely flaring by a factor of 2-4 in the outer disk
\citep{heyer2015,marasco2017}. Comparable ratios between the stellar and dust
scale heights are seen in other massive disk galaxies
\citep[e.g.,][]{xilouris1999}\footnote{Some high-latitude dusty clouds
  are seen in galaxies with high star formation rate intensities
  \citep[e.g.,][]{howk1999,rueff2013}, and some diffuse dust may be
  associated with the thicker atomic gas layer \citep[see arguments
    in][]{wild2011}, but the preponderance of clearly defined dust
  lanes \citep[80\% of edge-on SDSS galaxies;][]{holwerda2012} suggests
  the majority of dust in massive galaxies is indeed associated with a
  thin layer confined to the midplane.}.  When viewed at high spatial
resolution, the effective thickness of the cold dusty ISM may be even
smaller if the extinction along a line of sight is produced by
individual molecular clouds (which have sizes of $<\!50\pc$) within
the thicker dust layer.  As such, the treatment of the dust layer as
being infinitely thin is unlikely to strongly influence our results.

\subsection{Mapping the Fraction of Reddened Stars}

\begin{figure*}
\centerline{
\includegraphics[width=6.5in]{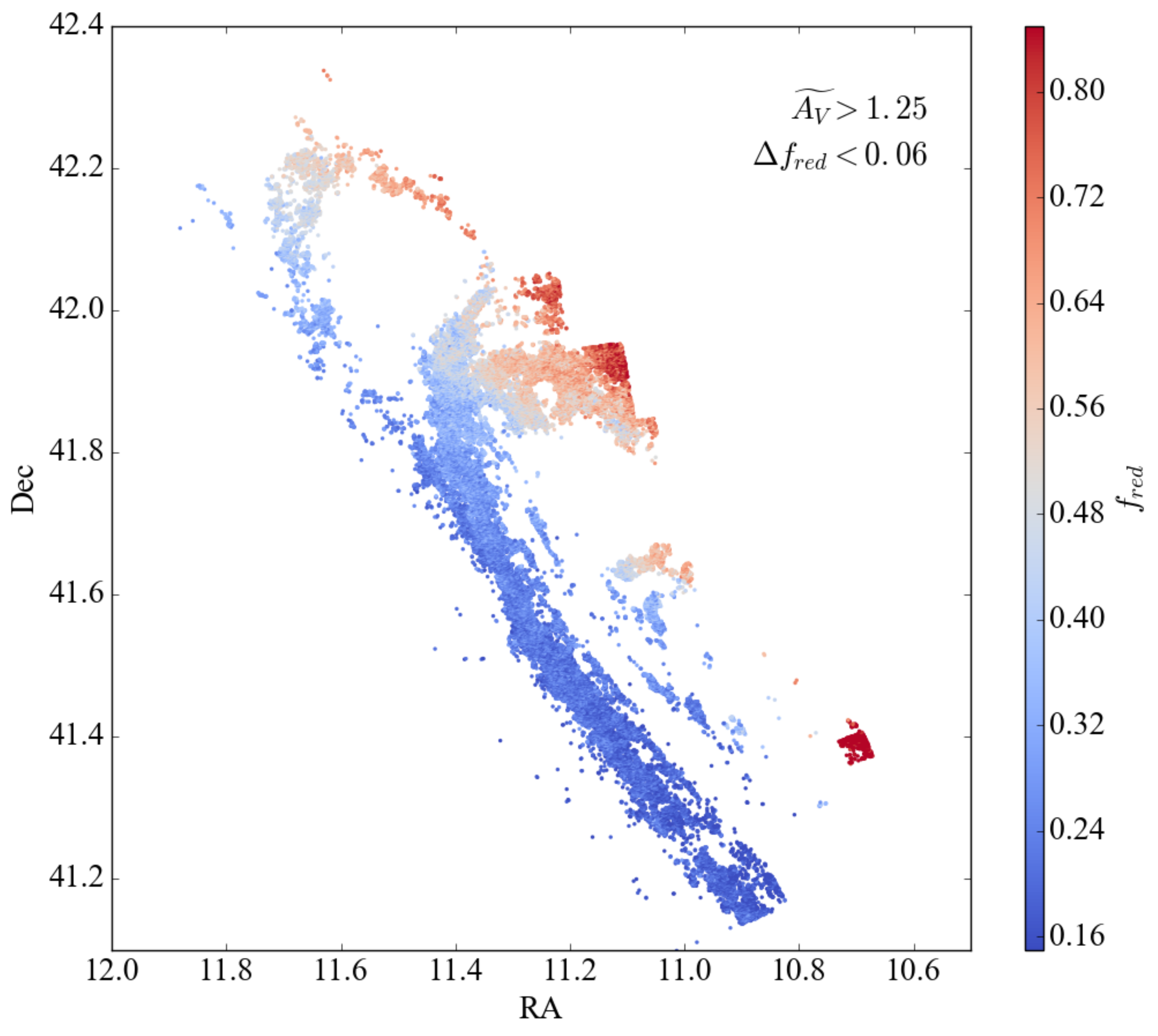}
}
\caption{Map of the fraction of reddened stars $f_{red}$, restricted
  to high extinction ($A_V>1.25\mags$) regions with well-measured
  values of $f_{red}$ ($\Delta f_{red} < 0.06$). There is a clear
  gradient in the fraction of reddened stars from the far side to the
  near side of the disk (i.e., left to right).  The majority of the
  PHAT survey area, which was initially targeted to avoid M31's dust
  lanes, has fewer than $\sim$25\% of its old stellar population
  behind the dusty ISM.  The center of M31 is in the lower right, at RA$\approx$10.68 and Dec$\approx$41.27.
\label{fredmapfig}\label{goodfredmapfig}}
\end{figure*}

Figure~\ref{fredmapfig} plots the full map of the fraction of
reddened stars \fred, restricted to points with well-measured values
of \fred\ ($A_V>1.25$ mag and $\Delta f_{red}<0.06$, where $\Delta
f_{red}$ is half of the difference between the 16\% and 84\%
percentile range for the MCMC-sampled posterior distribution of \fred, which
would be equal to the standard deviation for a Gaussian distribution).
In spite of the complexity of the extinction map presented in
\citet{dalcanton2015}, and the independent analysis of each plotted pixel, the fraction of reddened stars varies smoothly
and systematically across the disk.  As expected, the fraction of
reddened stars is 0.5 along the major axis, but diverges to much
smaller and larger values with increasing distance from the major
axis.  There is a steady shift in the fraction of reddened stars from
the near side of the disk (upper right) to the far side (lower left).
On the near side, the fraction of reddened stars is very high, as
would be expected from the optical morphology alone; this side of the
disk shows clear dust lanes, which just graze the edge of the PHAT
footprint.  In contrast, on the far side of the disk (where there are
no obvious strong dust lanes) the fraction of reddened stars is
extremely low, with fewer than $\sim$20\% of stars lying behind the
dust layer.

We note that the result in Figure~\ref{fredmapfig} is essentially by
design, given that PHAT specifically targeted the quadrant of M31 that
appeared the least affected by dust.  That said, the low reddening
fraction and the unobscured visual morphology does not actually
suggest that there is no dust on the far side of the galaxy
\citep[e.g.,][]{elmegreen1999}, given that the extinction maps clearly
show ample dust in this quadrant, as does mid- and far-IR emission \citep[e.g.][]{draine2013}.  Instead, the weakness of visible
dust obscuration on the far side of the disk is the result of the geometrical effect we are exploiting
in this paper.

The large observed range in \fred\ immediately suggests that the disk
of RGB stars is not thin. If it were, then there would be only a
modest radial range sampled along all lines of sight, and thus comparable
number of stars behind and in front of the thin dusty layer of the
cold ISM. In such a case, \fred\ would not vary dramatically from 0.5,
in sharp contrast to what we see here.  In retrospect, the fact that
even ground-based optical images of M31 show much stronger reddening
on one half of the galaxy is strong evidence for a thick disk as well,
although not as quantitatively useful.

\subsection{The 3-Dimensional Structure of M31's Stellar Disk} \label{fredsec}

To derive the disk thickness from the reddening map in
Figure~\ref{fredmapfig}, we return to the distribution of maximum
deviations from \fred, plotted in Figure~\ref{fredmaxfig} as a
function of inclination and $h_z/h_r$.  The observed extremes of
\fred\ identify a particular locus in the plot of inclination and
$h_z/h_r$.  If we then constrain M31's inclination from its observed
axial ratio (Figure~\ref{axialratiofig}), we can derive the value of
$h_z/h_r$ for M31's stellar disk.

\begin{figure}
\centerline{
\includegraphics[width=3.5in]{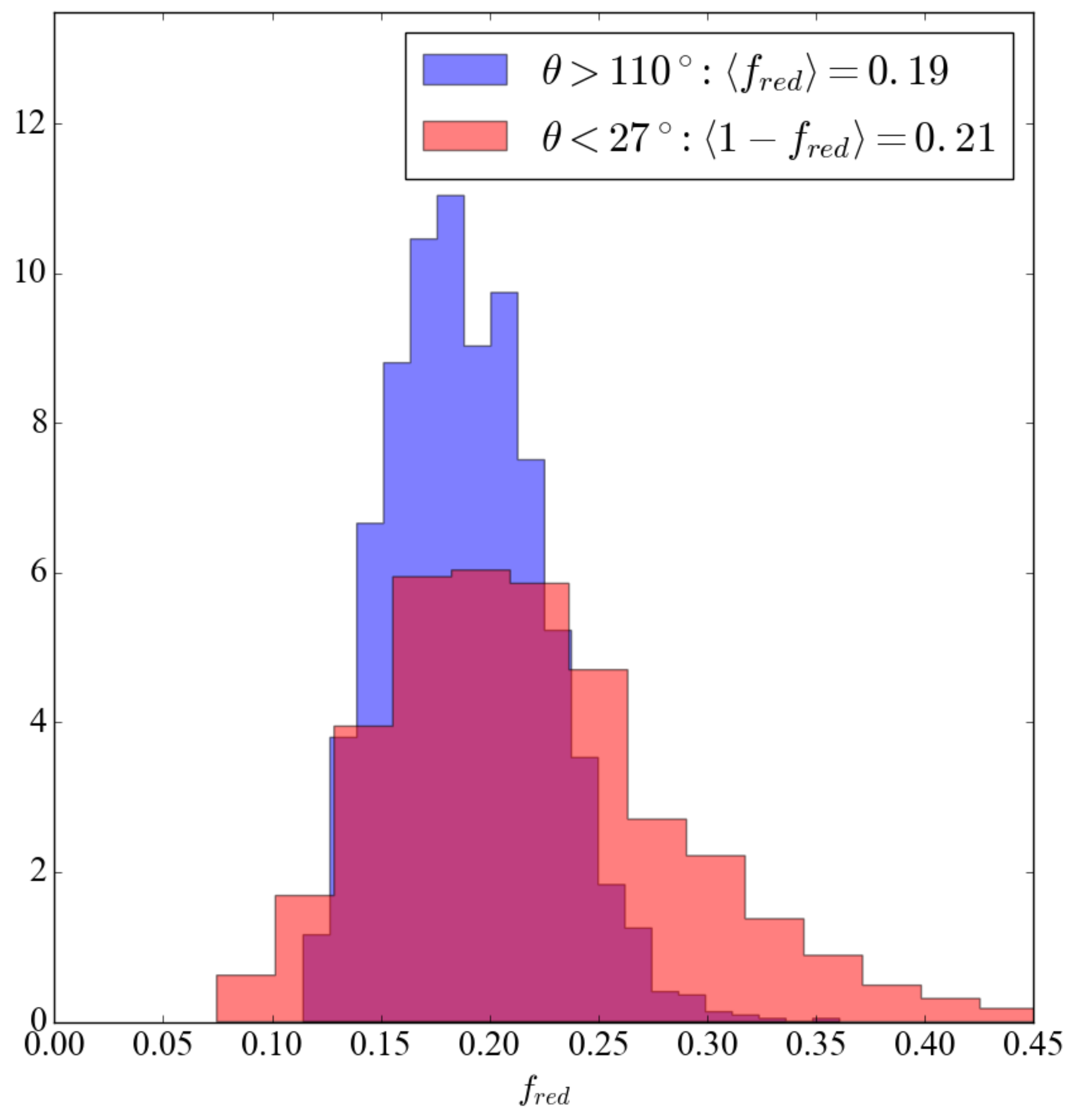}
}
\caption{The extremes of the fraction of reddened stars $f_{red}$,
  using same high quality measurements plotted in
  Figure~\ref{fredmapfig}. The minimum values of $f_{red}$ (blue
  histogram) are calculated in regions roughly along the minor axis
  ($\theta>110\degree$), and the maximum values (plotted as
  $1-f_{red}$; red histogram) are calculated at $\theta<27\degree$,
  due to lack of data along the minor axis for the highly reddened
  side of the disk. These maximum values are taken from much closer
  the major axis, and so are not expected to have reached their
  asymptotal values, unlike for the minimum values of $f_{red}$; this
  limitation is manifested as the
  tail to larger values of $1-f_{red}$ in the red histogram. In spite
  of the poorer sampling of regions with high $f_{red}$, the values of
  $f_{red}$ on the obscured, near side of the disk are in good
  agreement with expectations from the well-sampled far side, given
  that the mean $f_{red}(\theta>110\degree) \approx 1-f_{red}(\theta<27\degree)$,
  after culling the obvious tails at $f_{red}>0.27$ and $1-f_{red}>0.32$.
\label{meanfminfig}}
\end{figure}

We identify the most extreme observed values of \fred\ by plotting
histograms of \fred\ (blue) and $1-f_{red}$ (red) from the far and
near sides of the disk in Figure~\ref{meanfminfig}.
For both quantities, we plot a subset of the points from
Figure~\ref{fredmapfig}, after excluding regions at small deprojected
radii ($r<0.8^\circ$) and near the major axis, where we expect to find
less extreme values of \fred\ based on the models shown in
Figure~\ref{fredgridfig}.  On the far side of the disk (low \fred), we
selected points within $\pm20^\circ$ of the minor axis
($160^\circ<\theta<200^\circ$ where $\theta$ is the angular polar
coordinate for the deprojected disk; see the transformations to polar
coordinates described in the Appendix).  The near side of the disk is
not as well-sampled, and we were forced to expand our selection to a
wider range of angles ($\theta<50^\circ$) to include the region of
well-measured high values of \fred\ found in the $10\kpc$ star-forming
ring.  There are $\sim$1400 points in the far side subsample and
$\sim$400 in the near side sample, and most are drawn from between 2-3.5$h_r$, well into the regime where the value of \fred\ reaches its extremes (e.g., Figure~\ref{fredgridfig})

The histograms in Figure ~\ref{meanfminfig} are clearly peaked at
$f_{red} \approx 0.2$ on the far side and $f_{red}\approx0.8$ on the
near side, with very small tails towards $f_{red}=0.5$, and a central
width consistent with the uncertainties in \fred ($\left<\Delta
f_{red}\right>=0.043$ and $\left<\Delta f_{red}\right>=0.044$ for the
far and near sides respectively).  The tails are expected when
admitting points away from the minor axis, and are more significant
for the near-side data where we do not sample the large radius, minor
axis behavior of \fred\ as well.  We calculate robust means for each
distribution and find $f_{red}=0.214$ for the far side and
$1-f_{red}=0.180$ for the near side. 
If we
restrict the subset to even higher extinction regions ($A_V>1.75$~mag),
these values change to $f_{red}=0.212$ and $1-f_{red}=0.186$.  We have
also experimented with different choices of signal-to-noise cuts and
angular or radial extents, and find consistent answers with all
plausible choices.  The formal errors on the mean are quite small, but
it is reasonable to assume that the true uncertainties are larger and
potentially dominated by systematics, due to the small area sampled on
the near side, and the possibility that fits are biased differently
when $f_{red}$ is high versus when it is low.  As such, the difference
between the two sides of the disk are likely to better reflect the
underlying uncertainty, and we therefore adopt $0.303\pm0.02$ as the
maximum observed deviation from $f_{red}=0.5$, where the uncertainty reflects the likely amplitude of systematic effects, as indicated by the differences between the far and near sides.

\begin{figure}
\centerline{
\includegraphics[width=3.75in]{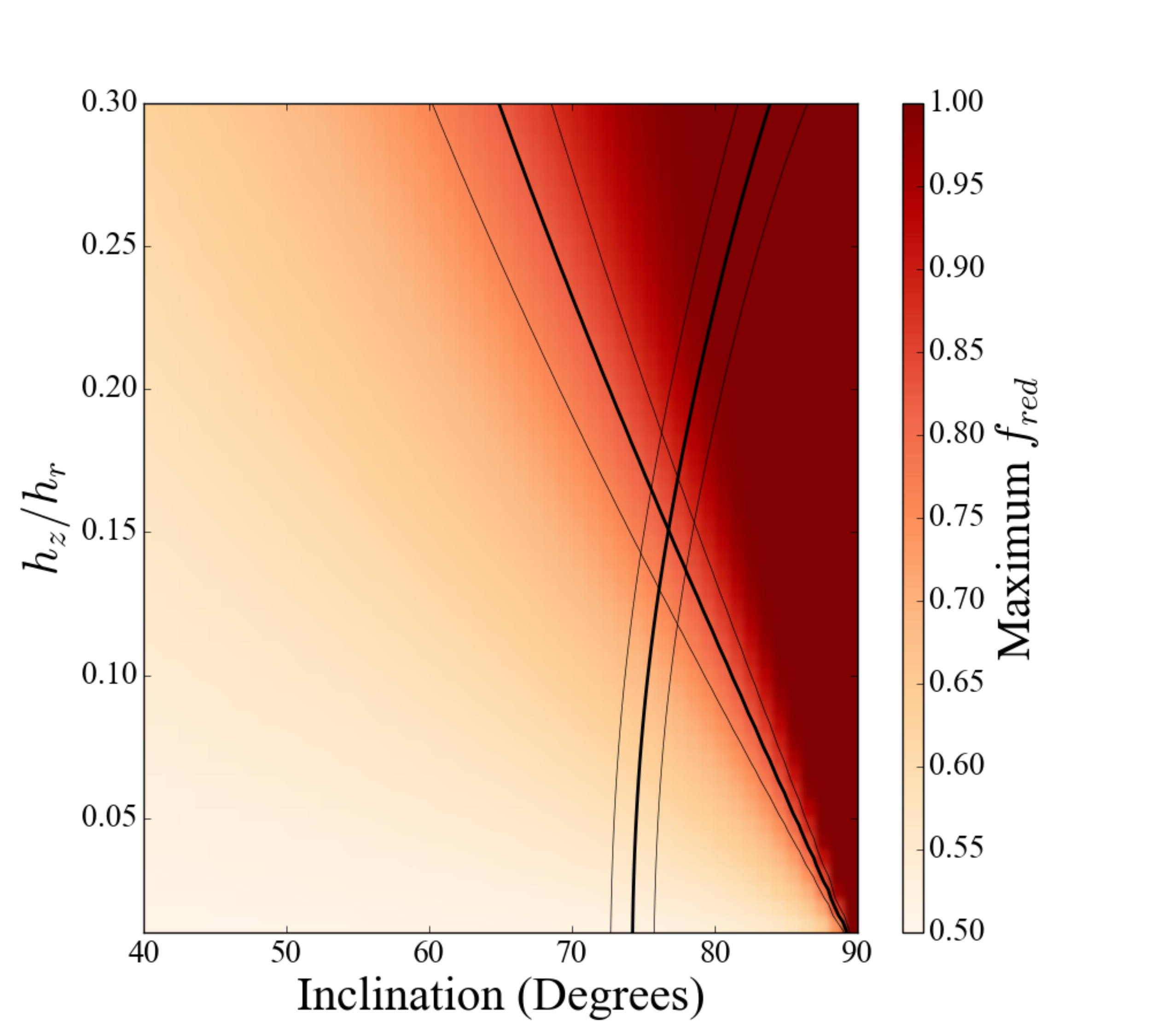}
}
\caption{Observational constraints on the maximum reddening fraction
  $f_{red}$ and the observed axial ratio for model inclined, thickened
  disks. The background color indicates the maximum reddenning
  fraction as shown in Figure~\ref{fredmaxfig} (i.e., measured along
  the minor axis at the surface brightness corresponding to $r=3h_r$
  for a face-on disk). One set of contours indicates the typical
  maximum value of $f_{red}=0.803\pm0.02$ observed in M31 (equivalent to
  the minimum values of $f_{red}=0.197\pm0.02$; see
  Figure~\ref{meanfminfig}). The other set of contours indicates the
  range of measured axial ratios from \citet[][;
    $b/a=0.27\pm0.01$]{courteau2011}, derived from their reported
  ellipticities ($\epsilon=0.73\pm0.01$), corresponding an inclination
  of $i=74\degree$ for an infinitely thin disk.  The
  intersection of these two independent measurements favors a typical
  disk thickness in the range of $h_z/h_r \approx 0.15$, and a larger
  true inclination of $i\approx77\degree$, which is in better
  agreement with the somewhat larger estimates of inclination from
  tilted ring fitting of M31's H{\sc i} velocity field
  \citep{corbelli2010,chemin2009}.
\label{fredaxialratiofig}}
\end{figure}

We plot the locus corresponding to this deviation
($f_{red,max}=0.803\pm0.02$) in Figure~\ref{fredaxialratiofig}, as a
function of inclination and $h_z/h_r$.  We now solve for $h_z/h_r$ by
constraining the inclination, using measurements of M31's disk
ellipticity ($\epsilon = 1 - b/a$).  A thorough analysis of 1- and
2-dimensional surface brightness fitting by \citet{courteau2011} finds
that $\epsilon=0.73\pm0.01$ for M31's old stellar disk, which
corresponds to $b/a=0.27\pm0.01$.  Joint kinematic plus photometric
decomposition in the $I$-band by \citet{dorman2013} find
$\epsilon=0.725\pm0.005$, in excellent agreement with
\citet{courteau2011}.  We adopt the more generous
$\Delta\epsilon=\pm0.01$ range from \citet{courteau2011} to be
conservative in our final estimate of the uncertainty on $h_z/h_r$.
In Figure~\ref{fredaxialratiofig} we reproduce the loci that traces
how this range of $b/a$ depends on inclination and $h_z/h_r$, based on
earlier results in Figure~\ref{axialratiofig}.

The intersection of the locii for the observed maximum $f_{red}$ and
the disk axial ratio places a tight joint constraint on the
inclination and thickness of M31's disk.  The inclination is
$77^\circ\pm0.5^\circ$, which is excellent agreement with values derived
from tilted ring fitting of M31's H{\sc i} velocity field over the
radii where we have measured $f_{red}$
\citep{corbelli2010,chemin2009}.

Figure~\ref{fredaxialratiofig} also places good constraints on the
vertical scale height $h_z$ relative to the radial scale length $h_r$.
We find $h_z/h_r = 0.14\pm0.015$, indicating the scale length of the
disk is between 6.5 and 8 times larger than the scale height.  This
ratio applies only to the older stellar disk, since \fred\ was
measured from near-infrared bright RGB stars.  It also does not
differentiate between any possible kinematic and/or metallicity
sub-components among the RGB stars themselves (such as the thin and
thick components identified by \citet{collins2011} or the ``kicked
up'' disk stars identified by \citet{dorman2013}), and instead reflects the properties
of whatever sub-population dominates the RGB stars.

We can convert the measured value of $h_z/h_r$ into the actual scale
height $h_z$ by using measurements of $h_r$.  We summarize some of the
most relevant existing measurements of $h_r$ in Table~\ref{hrtable},
focusing on analyses at longer wavelengths where the light is more
likely to be dominated by the RGB stars used in the measurement of
\fred.  We also include measurements of $h_r$ that directly analyze
the density of RGB stars \citep[e.g.,][]{choi2016}.  Although some individual papers quote
high-precision measurements for $h_r$, the true accuracy is much
poorer, largely because of the complexity of M31's true structure,
which features multiple central spheroids \citep[e.g.,][]{beaton2007},
bars \citep[e.g.,][]{athanassoula2006,choi2016,opitsch2017,blanadiaz2017}, and
an overdensity of light and stars at the $10\kpc$ ring
\citep[e.g.,][]{courteau2011,dalcanton2012}.  We adopt
$h_r=5.5\pm0.5\kpc$ as a realistic estimate of the appropriate scale
length and its uncertainty.  With this, we find

\begin{equation}
  h_z = 0.77\pm0.08\kpc \left(\frac{h_r}{5.5\kpc}\right).
\end{equation}

\noindent If we consider the uncertainties on $h_z/h_r$ and $h_r$ as
ranges allowed by systematic errors, then we expect $h_z$ to fall in the
range $625 - 930\pc$.  We note that this is substantially larger than
the likely scale height of the cold dusty gas, supporting the validity of
treating the dust as a thin screen within the stellar disk.

\begin{deluxetable*}{Cccp{8cm}}
  \tablecaption{M31 Disk Exponential Scale Length Measurements at Long Wavelengths \label{hrtable}}
  \tablehead{
    \colhead{$h_r$ (kpc)} &
    \colhead{Bandpass\tablenotemark{a}} &
    \colhead{Reference} &
    \colhead{Notes}}
  \startdata
    5.3\pm0.5  & 3.6$\mu$m & \citet{courteau2011} & Variety of methods including 2-D bulge$+$disk fitting and 1-D fits to major \& minor axis wedges at fixed P.A.\\
    5.91\pm0.27 & 3.6$\mu$ & \citet{seigar2008} & Fitting 1-D profile from ellipse-fitting surface brightness profile \\
    5.09-5.91 & 3.6$\mu$ & \citet{seigar2008} & Rotation curve fitting with different models \\
    5.76\pm0.1 & $I$ & \citet{dorman2013} & Joint kinematic$+$photometric modeling of velocities$+$image\\
    5.26\pm0.01 & $W1$ & \citet{choi2016} & 2-D bulge+disk fitting of {\emph{WISE}} mosaic \\
    5.56\pm0.45 & & \citet{choi2016} & 2-D bulge+bar+disk fitting of RGB star counts from PHAT data \\
    5.0-5.5 & \nodata & \citet{williams2017} & Stellar mass profiles derived from HST optical$+$NIR CMDs \\
    5.1\pm0.1   & \nodata & \citet{ibata2005} & RGB star counts in the outer disk (20--40$\kpc$)\\
  \enddata
  \tablecomments{All scale lengths calculated at a distance $785\kpc$.}
  \tablenotetext{a}{3.6$\mu$m images taken from {\emph{Spitzer}} IRAC images by \citet{barmby2006} and $I$-band images taken from \citet{choi2002} unless otherwise noted.}
\end{deluxetable*}

\section{The Thickness of M31's Stellar Disk in Context} \label{contextsec}

The measurements presented in Section~\ref{fredsec} indicate
that M31's stellar disk is moderately puffy ($h_r/h_z\approx7.1$) and
thick, with half of the stellar disk mass lying more than
$\pm0.53\pm0.05\kpc$ above or below of the midplane (assuming the half-height
$z_{1/2}=0.693h_z$ for an exponential disk, with $h_z = 0.77\kpc$).
We now place these measurements in the larger context of what is known
about the structure of disk populations in the Milky Way, and in
larger samples of edge-on galaxies.

\subsection{Comparison to the Milky Way} \label{MWsec}

Although the exact scale height of the Milky Way's disk remains
somewhat uncertain, scale heights as large as we measured for M31 have
never been reported for the Milky Way's thin disk.  The synthesis of Milky Way
structural measurements by \citet{blandhawthorn2016} argues for a
metal-rich ``thin'' disk component that is more than a factor of two
smaller than we find in M31 ($h_{z,thin}\approx300-350\pc$ from
photometric studies, or smaller values of
$h_{z,[Fe/H]>-0.3}\approx240-270\pc$ when subdividing populations on
metal abundance \citep[e.g.,][]{bovy2012a}).

\begin{figure*}
\centerline{
\includegraphics[width=3.25in]{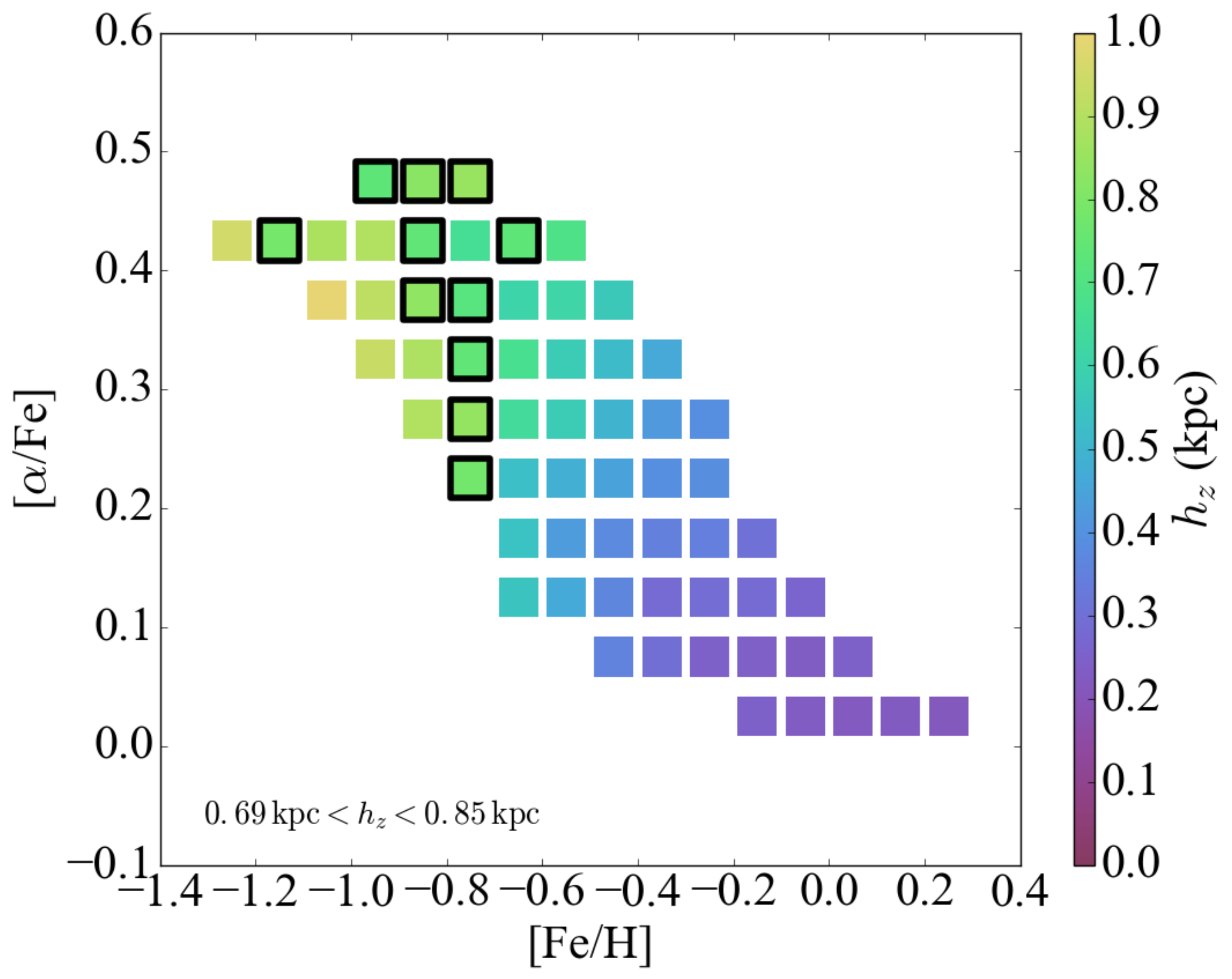}
\includegraphics[width=3.25in]{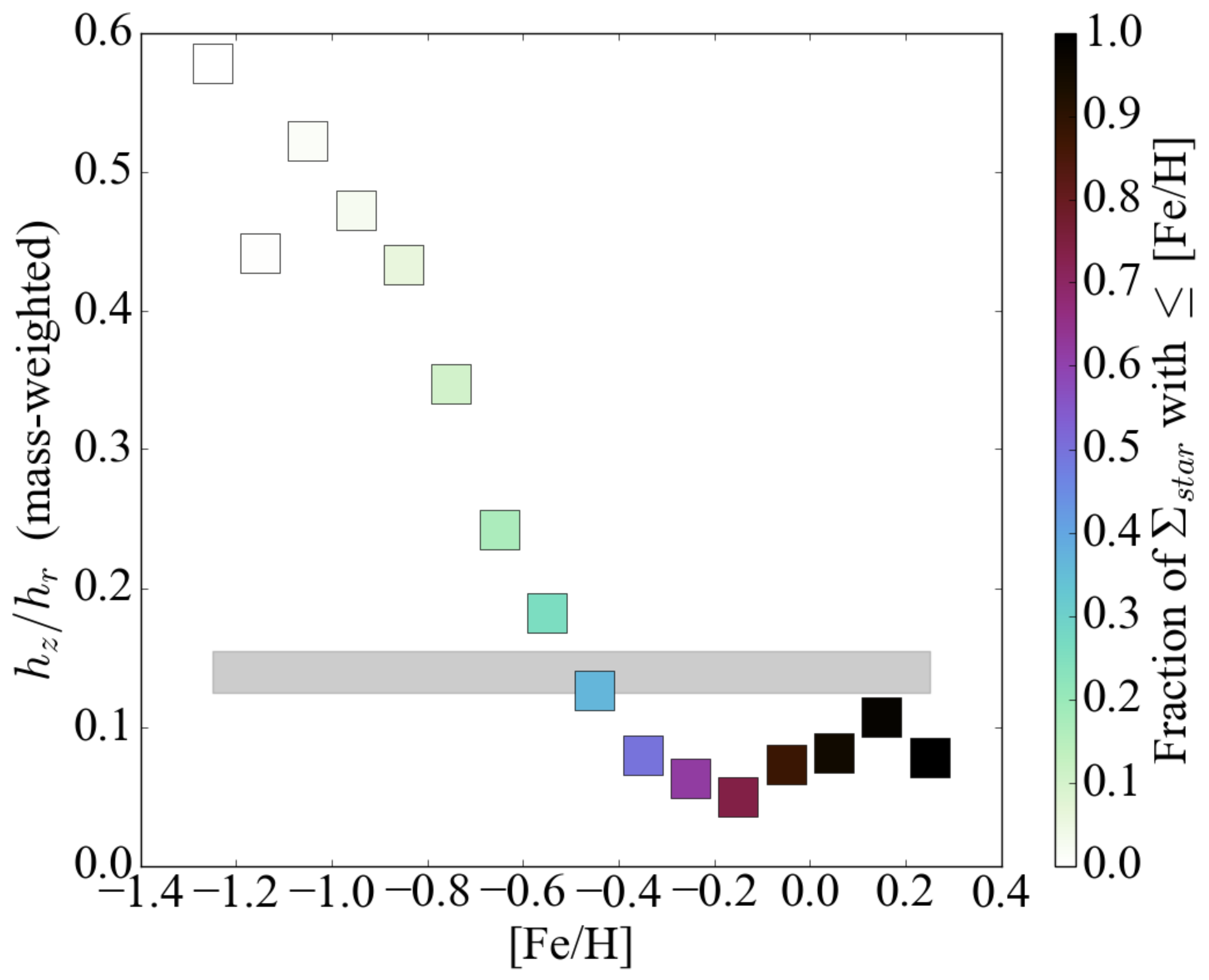}
}
\caption{Comparison between the structure of M31's RGB disk and the
  structure of the Milky Way's monoabundance populations from
  \citet{bovy2012}.  (Left) Monoabundance populations in [Fe/H] vs
        [$\alpha$/Fe], color-coded by their vertical scale height at
        the solar circle.  Populations whose scale height falls in the
        range allowed by the data in M31 are outlined in black.  Only
        low metallicity, $\alpha$-enhanced Milky Way populations have
        scale heights comparable to those we see in M31.  (Right) The
        axial ratio of the monoabundance subpopulations as a function
        of [Fe/H] calculated as a mass-weighted average of $h_z/h_r$
        over [$\alpha$/Fe] at fixed [Fe/H]. Points are color-coded by
        the fraction of the integrated local stellar density found in
        populations with metallicities equal to or lower than [Fe/H].
        The horizontal bar show the likely range of $h_z/h_r$ for
        M31's stellar disk.  Only low metallicity populations
        ([Fe/H]$\lesssim-0.5$) that make up $\sim$30\% of the Milky
        Way disk are as puffy as the M31 disk.
\label{bovyfig}}
\end{figure*}

In contrast, measurements of the scale height of the Milky Way's thick
and/or metal-poor disk are comparable to that seen for M31's RGB stars
\citep[$h_{z,thick} \approx 700-1200\pc$ or $h_{z,[Fe/H]<-0.25}
  \approx 690-770\pc$;][]{blandhawthorn2016,bovy2012a}.  We
demonstrate this correspondence in Figure~\ref{bovyfig}, where we plot
the vertical scale heights of the monoabundance populations from
\citet{bovy2012a} and outline the metallicity subpopulations that have
thicknesses consistent with that seen in M31. Only by looking at
metal-poor subpopulations with [Fe/H]$\lesssim-0.6$ and
$\alpha$-enchancements of [$\alpha$/Fe]$\gtrsim0.2$ do we find Milky Way
subpopulations with comparably thick disks.
The larger thickness of M31's stellar disk persists even if we
consider the possibility that we have overestimated $h_z$ by
multiplying the measured value of $h_z/h_r$ by too large a value of
$h_r$.  Among modern measurements, there are no disk scale lengths
smaller than $h_r\approx5\kpc$ (see Table~\ref{hrtable}).  If we adopt
this extreme value of the scale length, it would only drop our
estimate of $h_z$ to $700\pc$, which is still more comparable to the
Milky Way's modest thick disk than its dominant thin disk.

For an alternative comparison, we can avoid the complication of
scaling by an uncertain value of $h_r$ by instead directly comparing
our measured value of $h_z/h_r$ to the same ratio in the Milky Way
(although the latter is affected by significant uncertainties in the
Milky Way's $h_r$).  The current best estimates of $h_z=300\pm50\pc$
and $h_r=2.6\pm0.5\kpc$ from \citet{blandhawthorn2016}'s review give
$h_z/h_r=0.11^{+0.05}_{-0.03}$ for the Milky Way's dominant thin disk,
which is more than 30\% smaller than the dominant population in M31.
We can also compare to the axial ratios of the monoabundance populations in
\citet{bovy2012}, as shown in Figure~\ref{bovyfig}.  Their thin-disk
analog subpopulation (high metallicity, solar $\alpha$-enhancement,
which dominates the Milky Way stellar disk) has typical thicknesses
of $h_z/h_r=0.064-0.071$, which is twice as flattened as we have
measured in M31.  At the other extreme, their thick disk analog (low
metallicity, $\alpha$-enhanced) subpopulation has intrinsic
thicknesses of $h_z/h_r=0.31-0.43$, which is consistent with measurements from RR Lyrae stars \citep{mateu2018} in the Milky Way, but is more than twice as puffy
as M31's dominant stellar disk.  Thus, while M31's disk has a
comparable absolute scale height to the Milky Way's metal-poor thick
disk sub-population, its proportions are somewhat more disky, although
not nearly as flattened as the Milky Way's dominant metal-rich stellar
disk.

We note that unlike in M31, the Milky Way's thicker components only
make up a small fraction of its surface density at the Solar
circle. \citet{blandhawthorn2016} argue for $f_{thick,MW} \approx 12\%
\pm 4\%$, based on the available literature considered in their review.  A slightly
higher fraction of thick disks stars is suggested by the monoabundance
populations from \citet{bovy2012a} ($\lesssim30$\% of the stellar
mass; see Figure~\ref{bovyfig}), but the thick disk fraction is always
well below 50\% of the disk mass.  In contrast, our measurement of the
structure of M31's disk must reflect the stellar populations that
dominate the RGB, and would therefore be insensitive to a trace
population that only made up a small fraction of the RGB stars.  The
most direct comparison can be made by ignoring any subdivision of the
Milky Way into different disk components; when doing so,
\citet{bovy2013} find that the integrated stellar disk has a
exponential scale height of $h_{z,MW}\approx400\pc$ (based on the
total stellar mass distribution calculated in \citet{bovy2012}), which
again is close to a factor of two smaller than what we find in M31.

In summary, by multiple measures, M31's older stellar disk is
substantially thicker than the disk of the Milky Way, in spite of the
fact that it has an age and metallicity comparable to the Milky Way's
thin disk (mean age of $\sim\!4\Gyr$ and metallicities in the range of
[Fe/H]$\approx0$ in the inner disk and [Fe/H]$\approx-0.3$ in the
outer disk based on work by \citet{dorman2015} and
\citet{gregersen2015}).

\subsection{Comparison to other galaxies} \label{comparisonsec}

In addition to comparing to the Milky Way, we can compare our
measurement of $h_z/h_r$ for M31's RGB stars to structural
measurements of edge-on disks.  These measurements are necessarily
more uncertain than measurements in the Milky Way, due to the strong
effect of dust opacity.  Dust in the midplane of edge-on galaxies will
typically be optically thick, such that the measured surface
brightness profiles do not reflect the true stellar density.  Even ignoring the
effects of dust, typical extragalactic observations frequently lack
the spatial resolution to accurately fit the shape of the vertical
surface brightness distribution near the mid-plane, unless the
galaxies are very close.  Finally, there is also evidence that
the widely-used practice of fitting the observed 2d profiles yields
systematically different results than fitting true 3d models
\citep[e.g.,][]{bizyaev2014}. We attempt to minimize these issues by
focusing on the subset of studies that use NIR observations of very
nearby galaxies, but recognize that we are unlikely to have eliminated
the significant uncertainties, making the comparison with the Milky
Way probably the most direct analog of the measurements made in
M31.

We first compare our results to structural parameter fits of
$K_s$-band images of edge-on bulgeless galaxies in
\citet{yoachim2006}.  Although \citet{yoachim2006} eventually
decompose galaxies into thick and thin components, we focus on the
fits to models of a single edge-on double exponential disk, which
offer the best analog of the measurement we have made in M31, since both
measurements are dominated by structure of stars on the RGB.

\begin{figure}
\centerline{
\includegraphics[width=3.55in]{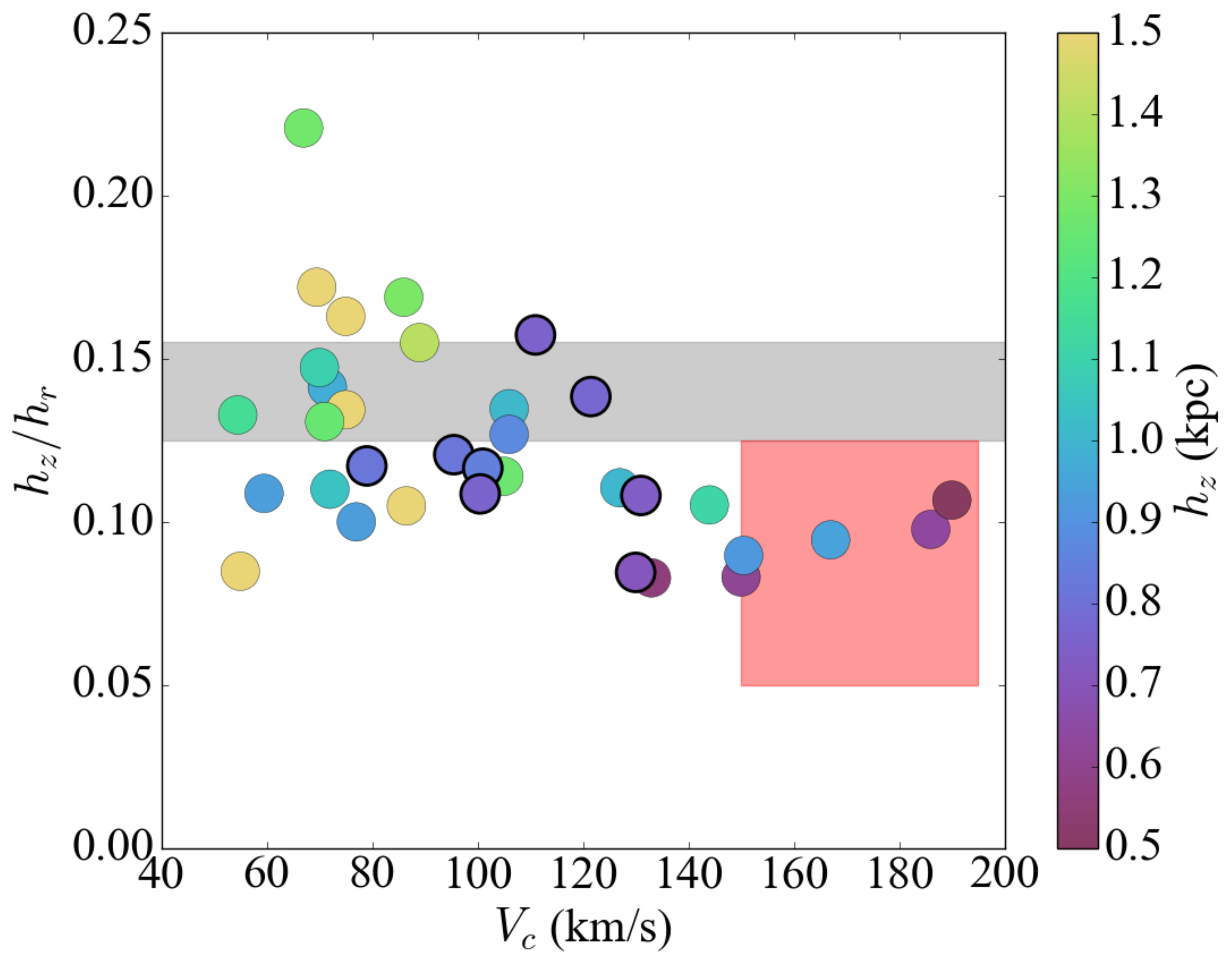}
}
\caption{Comparison between the structure of M31's RGB disk and the
  structure of edge-on disks from observations in the NIR.  Circular
  points are disk axial ratios ($h_z/h_r$) as a function of galaxy
  circular velocity, color-coded by scale height for the single-disk
  fits from \citet{yoachim2006}, based on fitting 2-dimensional models
  to $K_s$ imaging data for bulgeless galaxies selected from
  \citet{karachentsev1993}'s Flat Galaxy Catalog; points with heavy
  black outlines have values of $h_z$ within range of the values
  compatible with our results for M31.  The red rectangular region
  indicates the range of $h_z/h_r$ found for 2-dimensional fits of
  either 2MASS or Spitzer imaging of edge-on galaxies with
  bulge-to-total ratios greater than zero by \citet{mosenkov2015}. The vertical extent of this red region is also indicative of typical values for the Milky Way (e.g., Figure~\ref{bovyfig}). 
  The shaded grey region indicates the range compatible with
  observations in M31.  The massive ($V_c>120\kms$) bulgeless galaxies
  from \citet{yoachim2006} have structural parameters consistent with
  the Milky Way's dominant metal-rich disk (Figure~\ref{bovyfig}) but
  are systematically flatter than what we observe in M31. The disks of
  galaxies with bulges from \citet{mosenkov2015} are likewise better
  analogs of the Milky Way than M31.
\label{yoachimfig}}
\end{figure}

In Figure~\ref{yoachimfig} we plot the ratio of scale height to scale
length ($h_z/h_r$), as a function of galaxy circular velocity,
color-coded by the vertical scale height in kiloparsecs.  The range of
$h_z/h_r$ consistent with the observations in M31 (grey bar in
Figure~\ref{yoachimfig}) is noticeably larger than the
\citet{yoachim2006} measurements for massive disks ($V_c>120\kms$),
suggesting that M31's stellar disk is puffier than typical massive
disks in very late-type galaxies.  Although these galaxies may not be
ideal analogs for an Sb galaxy like Andromeda, they are consistent
with the properties of the Milky Way's dominant metal-rich disk
(Figure~\ref{bovyfig}).

We may find better analogs of M31 among the more diverse sample
analyzed by \citet{mosenkov2015}, also using 2-dimensional fitting
comparable to that in \citet{yoachim2006}, but for 2MASS $K_s$-band
and Spitzer 3.6$\mu$ imaging from the S$^4$G survey.  The red
rectangular region in Figure~\ref{yoachimfig} indicates the range of
$h_z/h_r$ found for galaxies with bulge-to-total luminosity ratios
greater than zero.  As with the bulgeless galaxies, M31 is again an
outlier in terms of its intrinsic axial ratio.

In total, the comparison of M31 to the measurements in
Figure~\ref{yoachimfig} suggest that M31's thickness is unusual
in the context of disk galaxies in general, not just compared to the Milky
Way.  

\subsection{Other Indicators of M31's Thickened Disk} \label{otherm31sec}

Our measurement of M31's disk is fully consistent with its other
observed features.  First, M31's stellar disk has a high velocity
dispersion, as would be needed to support a vertically extended disk.
In a series of papers \citet{dorman2012,dorman2013,dorman2015} used
Keck spectroscopy of stars selected in the PHAT footprint to measure a
high velocity dispersion for M31 RGB stars. They found a line-of-sight
velocity dispersion of $\sim\!90\kms$ over much of the disk beyond
$10\kpc$, rising to $120\kms$ in the inner disk at $5\kpc$. Large
velocity dispersions are also seen in the population of disk planetary
nebulae, both directly \citep[e.g.,][]{bhattacharya2019}, and as indicated by their significant
rotational lag due to asymmetric drift \citep{merrett2006}.

Unfortunately, the line-of-sight velocity dispersion is not a direct
measure of the vertical velocity dispersion, which is the component
that is most relevant to disk thickness. However, although the full
3-d velocity ellipsoid is unknown in M31, we can use the Milky Way as a
model to show that the vertical component is likely to also be high.  In
the Milky Way, the radial velocity dispersion is typically a factor of
1.4--2.1$\times$ larger than the vertical velocity dispersion
\citep[e.g.,][]{budenbender2015,sharma2021}, which would make the vertical
velocity dispersion $\sim\!43$--$57\kms$ if M31's velocity ellipsoid
were similar the Milky Way's, assuming the \citet{dorman2015}
line-of-sight measurement is dominated by the radial component.  For
comparison, the only component of the Milky Way with such a high
vertical velocity dispersion is the most metal-poor, $\alpha$-rich
subpopulation \citep[$\sigma_z\!\approx\!47\kms$, versus
  $\sigma_z\!\approx\!19\kms$ for the metal-rich, $\alpha$-poor
  disk;][]{budenbender2015}.

We note that an earlier paper by \citet{collins2011} also argued for a
thick disk in M31 on the basis of stellar kinematics, largely measured
in the outer disk ($\gtrsim\!15\kpc$, with the exception of one field
at $\sim\!10\kpc$).  They decomposed their measured velocity
distributions into a low velocity dispersion thin and high velocity
dispersion halo component, and then found evidence for an intermediate
dispersion rotating component that lagged the rotation of the thin
disk by $\sim50\kms$, which they then identified as a thick
disk. However, despite the similarity in nomenclature, it is unlikely
that the thick disk in \citet{collins2011} is strictly analogous to
the overall thickened disk we measure here.  The \citet{collins2011}
thick disk is a single subcomponent added to a dominant thin disk with
a $\sim\!36\kms$ line-of-sight velocity dispersion, whereas both we
and \citet{dorman2015} find that the majority of the old RGB stellar
population is suggestive of a hot, thickened component.

We suspect that a substantial part of the difference in interpretation
might be traced to the larger mean radius of \citet{collins2011}, and
to a lesser degree to the non-overlapping analysis regions (where PHAT
covers the northeast and \citet{collins2011} the southwest).  While
the majority of the \citet{collins2011} fields have $22-35\kms$ velocity dispersions for their thin disk component, their two innermost fields have thin disk
dispersions of $55-69\kms$, which is much closer to the values in
\citet{dorman2015} and consistent with the earlier $\sim\!50\kms$
velocity dispersions for the extended disk reported by
\citet{ibata2005}.  \citet{collins2011} also report that they had
difficulty isolating a clean thick disk component in the inner
regions, which would be consistent with \citet{dorman2015} and our
conclusion that the RGB stars in the inner disk are primarily in a
thick component.

Alternatively, some of the differences in the characteristic velocity
dispersion could be due to differences in methodology.
\citet{collins2011}'s velocity dispersions are based on decompositions
of the velocity histogram into multiple components using a Gaussian
mixture model and/or identifying the thick disk as those stars that
are not well-fit by a disk or halo component.  In contrast,
\citet{dorman2015} reports the weighted second moment of the entire
line-of-sight velocity distribution.  The former method (which was
needed to deal with the larger importance of the halo at large radii)
will always produce smaller velocity dispersions than treating the
entire distribution with a single component (which is well-justified
in the inner disk covered by Dorman et al (2015), as commented on in
\citet{dorman2012}).

We also considered whether the different conclusions about velocity
dispersions could be traced to differences in stars used in the analyses.
While \citet{dorman2015} analyze RGB and AGB
stars separately, \citet{collins2011} analyzes all the red stars
within a magnitude range that likely contains some degree of
contamination from younger AGB stars \citep[compare Figure 7 to Figure
  5 of][]{dorman2015}.  Although the strong age-dependent velocity
dispersion seen in \citet{dorman2015} could lead to some degree of
bias, we suspect this contamination is modest and unlikely to
produce the difference in velocity dispersion.

In addition to M31's thickened disk being compatible with the large
velocity dispersion measured in the PHAT footprint, it is also nicely
consistent with the merger history models recently proposed by
\citet{desouza2018} and \citet{hammer2018}.  In these models, M31 experienced a major $<$4:1 gas
rich merger $\sim\!2\Gyr$ ago, producing a high-velocity dispersion
stellar disk in the simulations of \citet{hammer2018}.  Because the \citet{hammer2018} model was tuned to in part reproduce the
high disk velocity dispersion measured by \citet{dorman2015}, its agreement
with the data is not surprising.  However, their simulations do show a
strong gradient in disk velocity dispersion, such that the outer disk
sampled in \citet{collins2011} would be lower velocity dispersion than
the inner regions sampled by \citet{dorman2015}. This difference may
well arise from the differing contributions of the two progenitors to
the two survey areas.  In the \citet{hammer2018} models, the outer
disk is dominated by stars pulled from the primary progenitor, whereas
the PHAT survey region contains a strong mixture of both the primary
and secondary progenitors.

Finally, there are tentative hints of a thickened disk in the three-dimensional chemical structure of M31, as inferred from ``made-to-measure" chemo-dynamical modeling of integral field spectroscopy of M31 \citep{gajda2021}.  The authors find flaring in the overal metallicity, and a tendency towards larger $\alpha$-enhancement at larger scale heights, which they argue is consistent with thickening due to a major merger as advocated for by \citet{hammer2018}, \citet{desouza2018}, and \citet{bhattacharya2019}.

\section{Implications}

The geometric confirmation that the majority of M31's stellar disk is
thick is perhaps not terribly surprising, given the violent merger
history visible in M31's extended halo, the high disk velocity
dispersion measured for its intermediate and old stellar populations,
and the strong evidence for a recent, global elevation of the star
formation rate 2$-3\Gyr$ ago \citep{williams2015,bernard2015}.  That
said, there are a number of interesting implications of having a
thickened stellar disk for M31, and for potentially identifying
thickened disks in other systems.

\subsection{Identifying candidate thickened disks}

The basic technique employed in this paper points towards an efficient
mechanism for finding equivalently thickened systems.  Inspection of
the models in Figure~\ref{fredgridfig} show that at fixed inclination,
the difference in apparent reddening on either side of the major axis
is an indicator of disk thickening.  Identifying moderately inclined
galaxies that show a dust lane on only one side would therefore be a
straightforward way to identify candidates for similarly thickened
disks.  

Inferring the exact value of disk thickening would require subsequent
modeling to calculate the actual reddening of dust lanes in the
imaging data, along with an estimate of the total dust column from
its emission at mid- and far-infrared wavelengths.   This approach was used in a prescient
paper\footnote{We gleefully note that this is also the first
  astronomical paper to use ``bogus'' in the title.} by
\citet{elmegreen1999} to explain the highly asymmetric $V-K$ profile
of the galaxy NGC~2841, and other similar galaxies like M64 (the
``evil eye'' galaxy NGC 4826) and NGC 3521.   For some applications, it
may be sufficient to only characterize one's lower-limits for
$h_z/h_r$, without the need for any additional modeling beyond setting
selection criteria. 

\subsection{Metallicity Gradients}

\begin{figure*}
\centerline{
\includegraphics[width=6.5in]{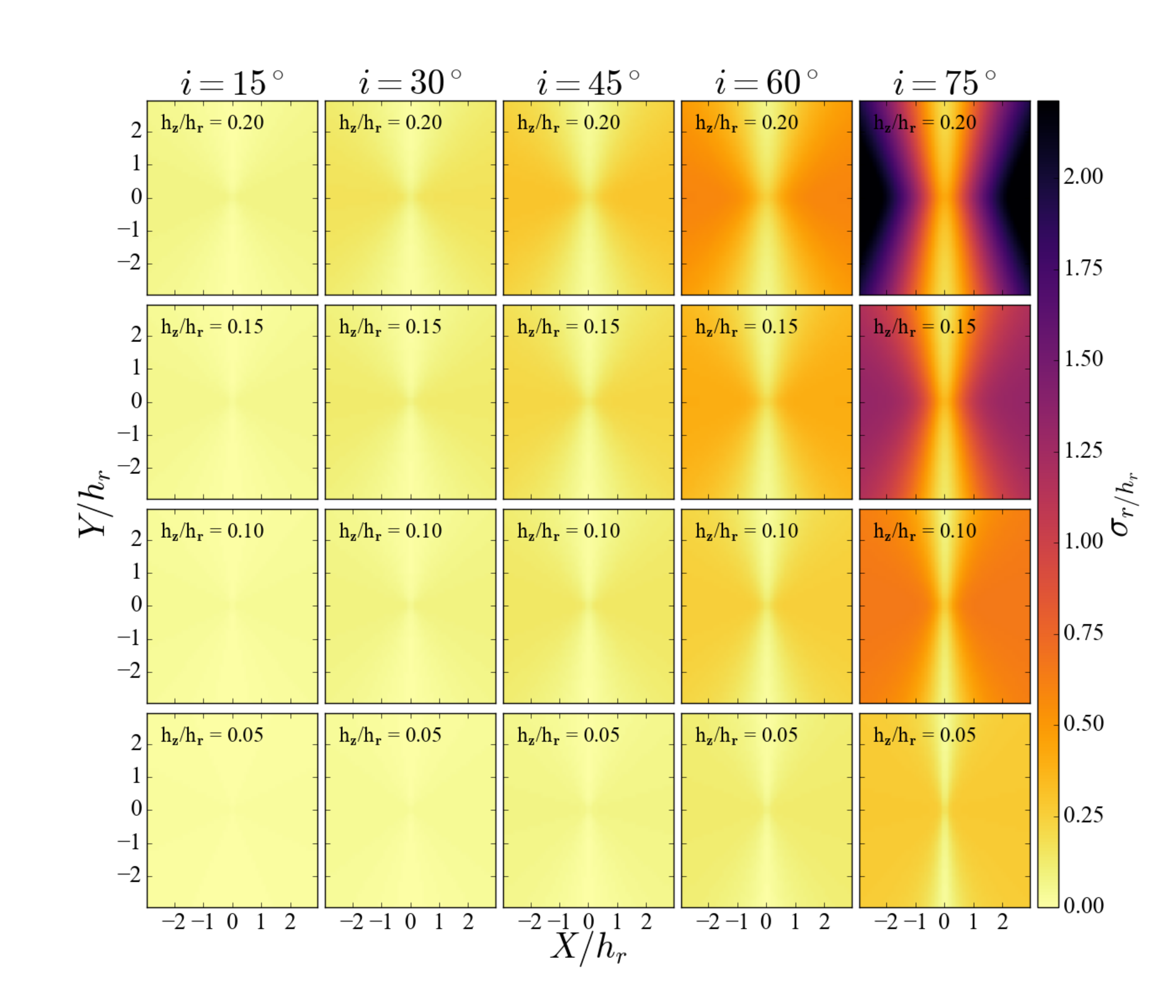}
}
\caption{Maps of the dispersion in radius along the line of sight, in
  units of the radial exponential scale length. The apparent
  inclinations of the model disks increases from left to right
  ($i=15\degree$, $30\degree$, $45\degree$, $60\degree$, \&
  $75\degree$), and the disk thickness increases from bottom to top
  ($h_z/h_r=0.05$, 0.1, 0.15, \& 0.2).  Away from the major axis, the
  range of radii probed increases with increasing inclination and disk
  thickness.  This effect will tend to wash out gradients in the measured
  properties of stellar populations, unless confined to along the major axis.
\label{radiusgridfig}}
\end{figure*}

M31's thickened stellar disk can potentially
wash out radial gradients in the disk.  In a thickened disk,
lines of sight away from the major axis probe a large range of radii, and as such, any
intrinsic gradients in the disk will tend to be averaged out. We show
the amplitude of this effect in Figure~\ref{radiusgridfig}, which
shows the mass-weighted dispersion in radii probed at each position in
an inclined, thick double exponential disk. For galaxies with
parameters like M31's (second panel from the top on the far right), a
single line of sight can sample stars from roughly $\pm1.5h_r$ in disk
radius.

The effect shown in Figure~\ref{radiusgridfig} may partially explain
the lack of any strong radial gradient seen in the planetary nebula (PNe)
metallicity distribution \citep[e.g.,][]{jacoby1999,kwitter2012,
  balick2013, balick2017} as well as its large scatter.  These
intermediate age \citep[$\gtrsim5\Gyr$; c.f.,][]{balick2017} stars are
likely to be distributed similarly to the RGB stars used in this
paper, and the long path length through the stellar disk will tend to
blend PNe from multiple radii into a single sightline, erasing
whatever intrinsic metallicity gradient survived M31's likely major
merger.  This merger-driven mixing
and the projection effects would also produce significant scatter, as
is evident in the PNe metallicity data down to radii of $\sim\!10\kpc$.

In contrast to the PNe, HII regions do appear to show a modest radial
metallicity gradient \citep[e.g.,][]{zurita2012}, which would be
consistent with their being in a thinner gas disk with negligible
projection effects.  Alternatively, gas flows
associated with a recent merger could potentially alter the
distribution of current gas phase metallicities, while leaving the
stellar metallicity distribution largely unaffected.

\subsection{A route to 3-D tomography}
The calculation of the reddened fraction maps in
Figure~\ref{fredgridfig} implicitly assumes that the disk has a
constant position and inclination angle, and that dust is in the
midplane of the stellar disk, everywhere.  These assumptions are likely
to be sound for an undisturbed, well-settled galaxy. For galaxies with
a more complicated recent interaction history, however, it is possible
for there to be global warps that manifest in the maps of
$f_{red}$, making the reddening maps a potentially powerful probe of the
3-dimensional galaxy structure.  Indeed, this effect has already been
demonstrated for the Large Magellanic Cloud, where \citet{choi2018} has modeled observed
reddening of red clump stars selected from the SMASH survey \citep{nidever2017} and revealed the presence of a warp in the far outer stellar disk.  Complementary work by \citet{yanchulova2021} in the Small Magellanic Cloud modeled the joint distribution of dust and stars from the red clump and RGB to infer the galaxy's line of sight depth and offset between the gas and stellar midplane.

Empirically, warps are primarily phenomena of the outer disk.
Stellar disk warps tend to emerge at $\sim\!R_{25}$ in edge-on
late-type disks, and are typically modest \citep[{$\sim\!1-5^\circ$}
  from the start of the warp; see][]{ann2006}\footnote{For optically
  thin outer disks, the actual warp radius is likely to be beyond the
  face-on value of $R_{25}$, due to projection}.  In the inner
regions, there is presumably enough stellar and dark matter mass to
keep the different components coupled together
\citep[e.g.,][]{ostriker1989,pranav2010}.  In practice, examining
warps in edge-on galaxies leads to observational biases that would
suppress detection of deviations from a perfectly flat aligned
disk in the inner disk, provided those deviations were modest and
obscured by the dust lane.  Indeed, low-amplitude warps or ``waves'' in the
stellar disk have been revealed at the solar radius (and beyond) in
the Milky Way \citep[e.g.,][]{widrow2012,xu2015,pricewhelan2015,
  ferguson2017,carrillo2018,schonrich2018,kawata2018, gaia2018, xiang2018, ramos2021, gaia2021, ding2021, laporte2022} and numerical
simulations \citep[e.g., recently][]{donghia2016,gomez2017,
  chequers2018,laporte2018, gomez2020, grionfiho2021, poggio2021, hunt2021}. 

Global warps can potentially be identified in reddening fraction maps.
HI studies have long used changes in position angle with radius (via
the kinematic line of nodes) to indicate the presence of warps
\citep[e.g.,][]{briggs1990}. Similarly, one could use the locus of
$f_{red}=0.5$ to define the position angle as a function of
radius. Inspection of Figure~\ref{fredmapfig} shows that 
$f_{red}=0.5$ primarily along a single axis.  However, there are
also small deviations of the order of a degree or two in localized
regions that could indicate warps.  The amplitude of these position angle deviations
are consistent with expectations from kinematical analysis of M31's
H{\sc i} distribution \citep{chemin2009, corbelli2010}.

Additional features could be produced in reddening fraction maps if
the gas layer that carries the dust becomes misaligned with respect to
the bulk of the stars. These misalignments can be due to the differing
dynamical responses of hot dissipationless (stellar) and cold
dissipational (gaseous) disks, or due to different intrinsic angular
momenta (say, if the gas had a different accretion origin).

Empirically, the gaseous and stellar disks appear to be globally
well-aligned (to within a few percent) in the vast majority of edge-on
galaxies, \citep[see discussion in][]{vanderkruit2011}.  This is not
unexpected in general, given that if both the gas and stars are
responding to a large-scale gravitational pertubation, then both will
experience the same potential and maintain their relative alignment if
given enough time to respond.  However, in simulations such as
\citet{gomez2017} it is clear that there are small-scale vertical
disturbances in the cold gas that are not reflected in the stellar
disk (compare their Figures 2 and 4, for example), most likely because
of the intinsic difficulty in producing small-scale pertubations in
high velocity dispersion dissipationless disks.  Comparable gas
deviations may have already been detected in the Milky Way
\citep[e.g.,][]{levine2006} and in M31 in the early work by
\citet{braun1991}.

Small-scale gas-star deviations could also potentially be traced by
the distribution of $f_{red}$.  If the gas is pulled away from the
midplane, the value of $f_{red}$ at that location will shift.  For
example, Figure~\ref{fredmapfig} shows a ``spur'' of gas at
$\alpha\approx11.3^\circ$, $\delta\approx42.0^\circ$ with a uniform
value of $f_{red}\approx0.5$ over a larger range in azimuth than
expected from the models in Figure~\ref{fredgridfig}.  It is possible
that this indicates a region where gas has been pulled away from the
midplane.  Similarly, accreted gas that has not yet settled to the midplane could also appear as a localized discrepancy in the reddening fraction.

In the future, it should be possible to create a full 3-D tomographic
model of M31 by combining the detailed structure of the reddening map
with a \citet{braun1991}-style analysis of modern high-resolution
H{\sc i} data.  Such a map could open up the possibility of studying
many effects of coupled gas and stellar dynamics in a system other
than the Milky Way.

\section{Conclusion}

``Dust geometry'' has long been a topic of interest, both for its impact on galaxies' light distribution \citep[e.g.,][]{disney1989,witt1992,calzetti2001} and for the study of the ISM itself \citep[e.g., see the review by][]{galliano2018}. This paper expands on the above work to highlight a less-widely appreciated role for measurements of dust geometry to be a potentially useful tool for understanding galaxy structure.  

We have explained the optical morphology of M31 as being shaped by the combined structure of the stars and the dusty ISM.  M31 shows a pronounced variation of the fraction of reddened RGB stars
from one side of the major axis to the other.  We measure this effect through modeling the NIR CMD, and find that on one side $\sim$80\%
of stars along the line of sight are reddened, and on the other as few
as $\sim$20\% of the stars are obscured by midplane dust.  This
variation can be straightforwardly explained by assuming that the path length
through the stars is long enough to probe a large range of galactic
radii.  In this model, path lengths where the inner galaxy is on the
near side of the dust will have more stars in front of
the dust, and thus a low fraction of reddened stars.

We have interpreted the quantitative measurement of the reddening fraction by generating models of a thin dust layer embedded in a stellar disk with varying viewing angles and ratios of the radial
to vertical exponential scale heights. By comparing the observed
amplitude of the reddening variation and the observed axial ratio of
M31 to these models, we constrain the old stellar disk of M31 to have
$h_z/h_r=0.14\pm0.015$.  This axial ratio implies that the vertical
exponential scale height is $770\pm80\pc$, for modern measurements of
M31's exponential disk at wavelengths dominated by RGB stars.

These measurements suggest that M31's inferred scale height and disk axial ratio has far more in
common with the Milky Way's thick disk than its thin disk, in spite of
being rather metal rich \citep[{[Fe/H] $\gtrsim
    -0.2$;}][]{gregersen2015,escala2023}.  It is likewise unusually thick when
compared to other edge-on galaxies.

While somewhat unusual, M31's thickened disk is fully compatible with the evidence that M31
has experienced a rather major merger in the past $2-3\Gyr$ \citep[see
  evidence compiled in][and \citet{desouza2018}]{hammer2018}. Mergers have long been
recognized as a means for vertically heating stellar disks \citep[e.g.,][]{quinn1993}, and once a stellar disk 
is heated, it is typically unable
to dynamically cool back into a colder, thinner disk.  M31's star
formation rate is currently relatively low and typical of a ``green
valley'' galaxy \citep{mutch2011}, and as such has not had time to
regrow a thinner disk of RGB stars.  We suggest that M31's morphology
of highly asymmetric reddening can be used as a generic criterion for
identifying comparable systems that are ``thick disk'' dominated.

We note that the large radial range probed within M31 is
likely to impact measurements of radial gradients of stellar
properties.  In general, any metallicity and age gradients associated
with $>2\Gyr$ old populations will tend to be washed out by projection
of multiple radii into a single projected position, making them appear
to be weaker than they truly are.  These effects are most significant away from the major axis, and would be expected in any ``puffy'' disk viewed at moderate inclination.

In addition to providing knowledge about M31 itself, the work here reinforces some of the key points of \citet{witt1992}'s \emph{cri de couer}, most notably that the amount of reddening can be largely decoupled from the amount of dust.  Both qualitatively and quantitively, M31 shows
a dramatic variation in the fraction of reddened stars, which obscures the large amount of dust on the southern half of the major axis.  This single case offers a cautionary tale about the need to have a flexible treatment of dust geometry when modeling or interpreting spectra or photometry of resolved galaxies, particularly when those galaxies are inclined or warped.

\begin{acknowledgements}
The authors wish to thank Jo Bovy for providing an electronic version
of the data used in Figure~\ref{bovyfig}, and Bruce Elmegreen for
pointing out his earlier work on explaining unusual reddening
distributions. JJD also thanks Adrian Price-Whelan, Vasily Belokurov, and Meredith Durbin
for support and input in the final stages of this work. JJD gratefully acknowledges the hospitality of the
Max-Planck Institut f\"ur Astronomie during part of this work. This
work was supported by the Space Telescope Science Institute through
GO-12055. The Flatiron Insitute is funded by the Simons Foundation. The authors made liberal use of Astropy, a
community-developed core Python package for Astronomy
\citep{astropy2013}, as well as numpy, scipy, and matplotlib
\citep{oliphant2007,hunter2007}.  This research has made use of NASA's
Astrophysics Data System Bibliographic Services and IPAC Science Archive.   

\end{acknowledgements}

\appendix

\section{Calculating Polar Coordinates for an Inclined Disk} \label{polarsec}

For the analysis in this paper, we map right ascension $\alpha$ and
declination $\delta$ into polar coordinates $r$ and $\theta$ on the
deprojected disk.  We assume that the disk is centered at ($\alpha_0$,
$\delta_0$), and has an inclination $i$ and a position angle $\phi$.
This disk will have an apparent axis ratio of $b/a = \cos{i}$ and
eccentricity $\epsilon = \sqrt{1 - (b/a)^2}$.  If we define angles relative
to the center of disk (within a tangent plane)
$\Delta\alpha=(\alpha - \alpha_0)\cos{\delta_0}$ and $\Delta\delta =
\delta - \delta_0$, then the deprojected radius is

\begin{equation}  \label{radiuseqn}
  r^2 = (\Delta\delta \cos{\phi} + \Delta\alpha \sin{\phi})^2 +
        \left[\frac{1}{\cos^2{i}}\right]\,
        (\Delta\delta \sin{\phi} - \Delta\alpha \cos{\phi})^2,
\end{equation}

\noindent and the angular polar coordinate is

\begin{equation} \label{thetaeqn}
  \cos{\theta} = \sqrt{1 - \left(\frac{y}{r}\right)^2}
\end{equation}

\noindent where

\begin{equation}  \label{yeqn}
  y = \Delta\delta \cos{\phi} + \Delta\alpha \sin{\phi}.
\end{equation}

These relations hold for either an infinitely thin disk or the
midplane of a thickened disk.


\bibliographystyle{aasjournal}  
\bibliography{references}
\end{document}